%% file: main.tex
\documentclass{article}

\include{macros}

\title{Measurements and linearized models for golf ball bounce.}
\author[$\dagger$,1]{Stanis{\l}aw W.~Biber}
\author[2]{Kristian M.~Jones} 
\author[1]{Alan R.~Champneys}
\author[1]{Riku Green}
\author[1]{Robert Szalai}
\affil[1]{Department of Engineering Mathematics, University of Bristol, BS8 
	1TW, UK} 
\affil[2]{Equipment Standards, R\&A Rules Ltd., St Andrews, Fife, UK}
\affil[$\dagger$]{Corresponding author: s.biber@bristol.ac.uk}

\date{\today}

\begin{document}

\maketitle

\begin{abstract}
A detailed set of experiments are described that capture over a 1000
different instances of the bounce of a golf ball. Video analysis is
used to capture velocity and spin immediately prior to and subsequent
to each bounce for a wide variety of landing conditions. Data are
presented from two different turfs; one artificial and one from a
typical tee.  Measurement errors and repeatability are analysed. The
data are compared to predictions from models of rigid bounce with
friction, including Penner's modification to account for
elasto-plasticity.  Coefficients of restitution and friction, and Penner's
effective contact angle are fit from the data. A better fit to the data is found using a non-physical piecewise-affine landing to lift-off
relationship, which distinguishes between cases that bounce in pure
slip from those that undergo rolling. Nevertheless, even balls that
undergo rolling are typically found to lift off slipping, having
undergone spin reversal. The findings suggest that further effort
needs to be spent on finding simple physics-based models of golf-ball
bounce.
\end{abstract}

\bigskip
\begin{center}
\end{center}

\section{Introduction}
After the impact between club face and ball, the golf shot can be
divided into two main components; the free-flight phase, and the bounce and roll phase. While ball flight is relatively well understood
\cite{quintavalla_flight}, there remains much uncertainty and a paucity of data against which to fit a model for the bounce of a golf ball. 

Elementary models of ball bounce are based on 
Newtonian restitution theory, which assumes am instantaneous, rigid bounce
during
which kinetic energy is dissipated by a constant factor $r^2$,
see e.g~\cite{Daish}. 
In many ball sports, such as tennis, football 
or cricket, and indeed for the launch of a golf ball from a club face,
such approaches can be improved by modelling an elastic ball that has finite contact time with a rigid surface, see e.g.~\cite{brake_analytical,cricket_impact,cordingley_phd,ghaednia_review}. 

For the interaction between golf ball and turf, however,
the ball remains relatively rigid whereas the turf exhibits elasto-plastic
behaviour. A simple modification to the rigid bounce theory to account for such interactions is that due to Penner \cite{penner2002} who argues that the elastoplastic forces during contact are equivalent to
the bounce against a rigid surface rotated 
through an angle $\beta$. Based on limited data, Penner proposed a general formula  for calculating such an angle, which depends linearly on the inbound angle and speed. Nonetheless, the physical rationale behind it would appear questionable.

Haake \cite{haake_apparatus} presented the largest data set on golf ball bounce
that we are aware of in the public domain. More than seven hundred
measurements were taken, including different in-bound conditions and at different golf courses. These data were used to examine the effectiveness of different viscoelastic models of golf ball bounce. 
A two-layer system of springs and dampers was found to
show best agreement with the experimental data, yet it was unable to
predict all features of lift-off accurately. Also, the two-layer model has many parameters, not all of which are physically amenable. 
Subsequent models have typically been evaluated against rather less
data, and are thus limited in their predictive power. 

To advance understanding of modern golf ball bounce and to motivate a new
generation of more accurate models (see, e.g.~\cite{biber_analysis}
which includes tangential as well as normal compliance), it has become
clear more accurate data is required. As a consequence, we have collected new data  on the  the motion of a golf ball, immediately before and after impact,
over a wide range of inbound speed, angle and spin conditions, using state-of-the art image analysis, not available at the time of Haake's work. We also attempt a systematic treatment of the effects of measurement error and variability.
The results are presented for two different ground types; an 
artificial and a natural turf. We shall also examine whether these data
are consistent with rigid-bounce models, either with or without Penner's
effective slope angle.

The rest of the paper is outlined as follows. Section 2 explains the methods of our study, including experimental set up, data analysis and how we fit the
models used for comparison purposes. Section 3 contains the results, which are discussed in Section 4.

\section{Methods}

\subsection{Experimental set up}

Two experimental campaigns were performed. In both cases a modern,
multi-piece golf ball, as used on the professional tours, was fired 
from a modified baseball launcher aimed downwards towards the designated surface. 
The ball was positioned in the launcher between two spinning wheels -- one at 
the top and one at the bottom. The spin speed of each wheel is controlled independently. 
Variations of these speeds (including possible reverse spin of the upper wheel) allowed a full range of realistic speeds and spins of the golf ball to be attained, and adjustment of the incline of the launcher allowed variation of the 
landing angle. The wheels of the launcher were aligned so that the axis of the spin imparted on the ball was perpendicular to its
plane of motion; thus avoiding any significant side-spin on the ball. 

During the first campaign (A) the ball was bounced off a
premium artificial golf teeing turf, 30 mm thick, adhered to a rigid
surface (wood with a thickness of 38 mm). In the second campaign
(B) the ball was bounced off a well-maintained teeing
area. In both campaigns, the launcher was moved between each repeat,
so that the ball hit a different, undamaged piece of the surface for
each recorded impact.

\begin{figure}[!b]
	\centering
	\begin{subfigure}{0.47\linewidth}
		\centering
		\begin{overpic}[width=0.9\textwidth]{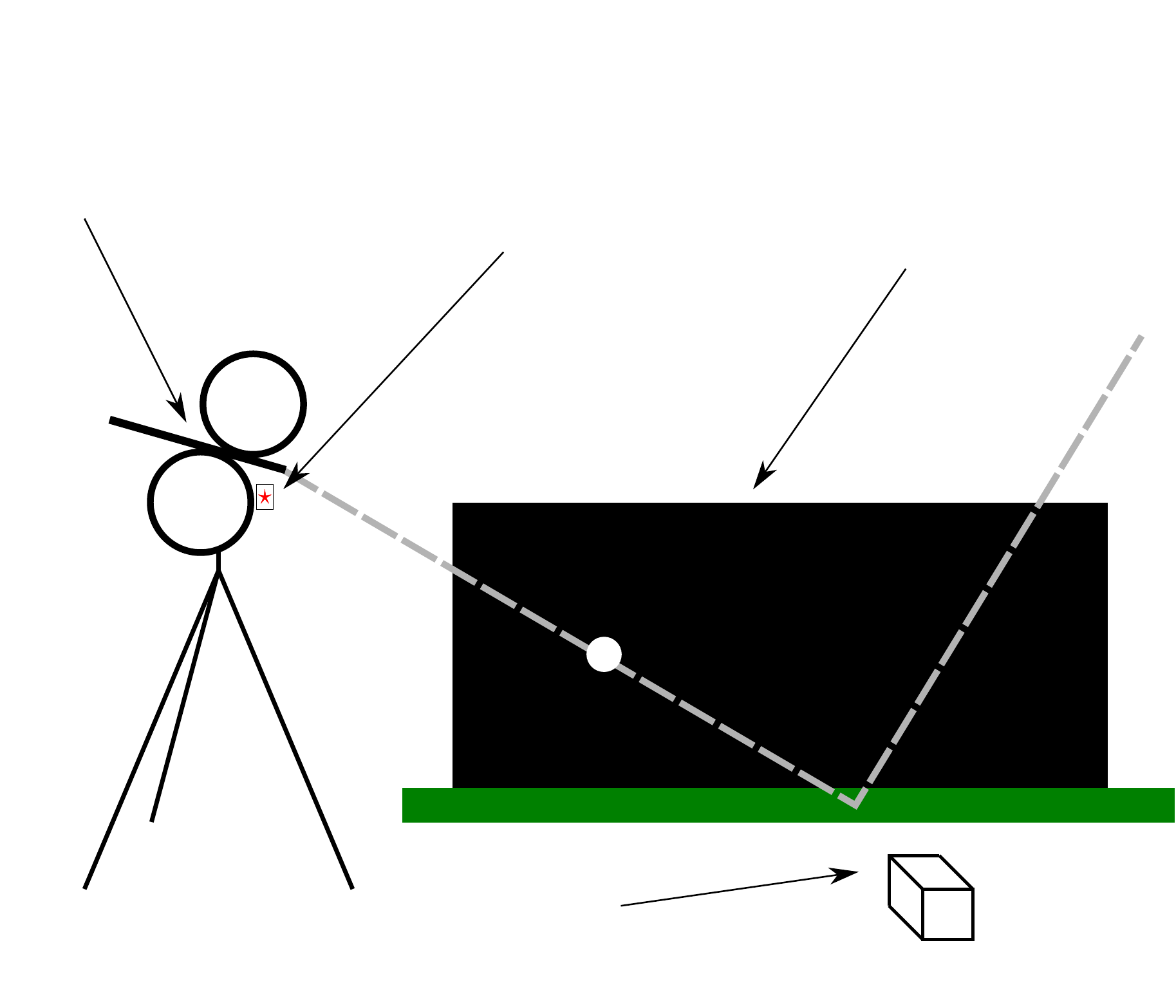}
			\put(25,66){Laser trigger}
			\put(56,65){Contrasting (black) background}
			\put(28,5){High speed camera}
			\put(0,70){Launcher}
		\end{overpic}
		\caption{ }
		\label{sfig:set_up_sketch}
	\end{subfigure}
	\begin{subfigure}{0.47\linewidth}
		\centering
		\includegraphics[width = 
		0.5\linewidth]{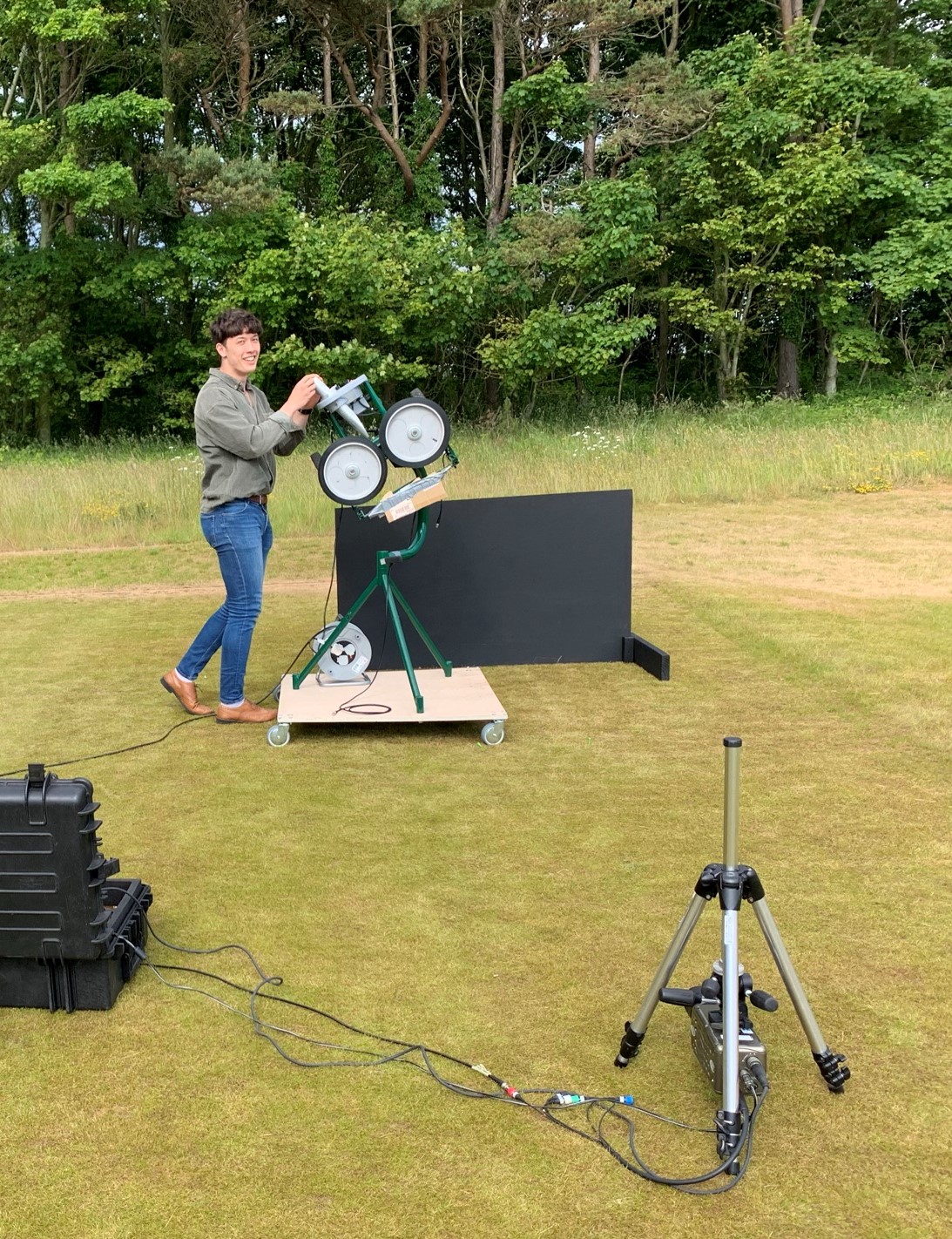}
		\caption{ }
		\label{sfig:set_up_photo}
	\end{subfigure}
	\caption{Experimental set up. The ball was launched from a modified 
	baseball launcher. On leaving the launcher, the ball triggered a laser 
	sensor which in turn activated a high-speed camera. (a) Annotated 
	schematic drawing of the setup. (b) Photograph from the experimental 
	session. }
	\label{fig:set_up_experiment}
\end{figure} 

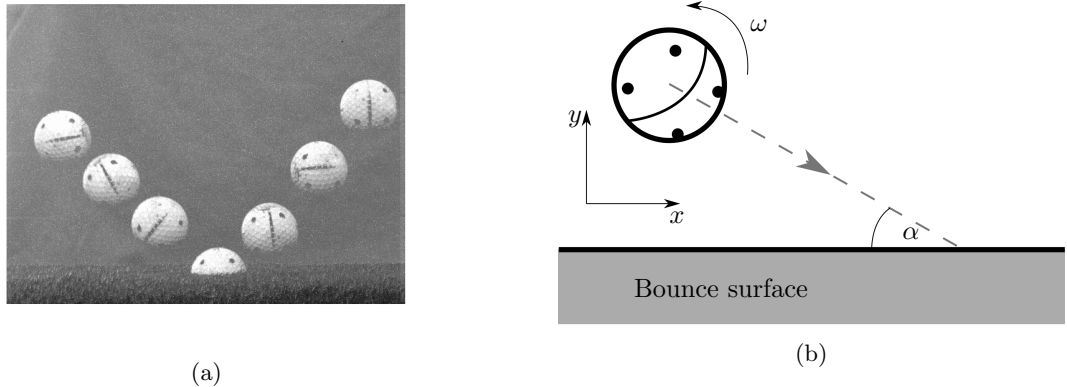
\begin{figure}[!htb]
	\centering
    \begin{subfigure}{0.48\linewidth}
        \centering
        \includegraphics[width = 0.85\linewidth]{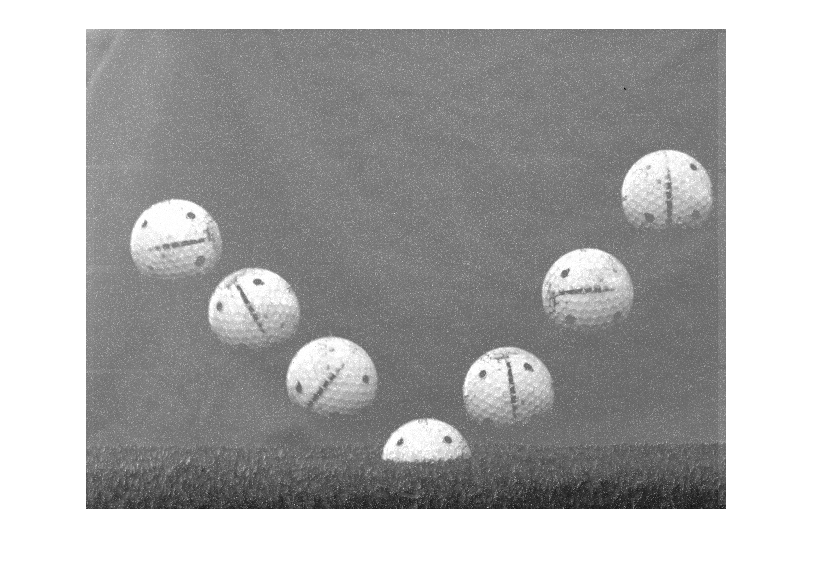}
        \caption{ }
        \label{fig:ball_trace}
    \end{subfigure}
	\begin{subfigure}{0.48\linewidth}
	    \centering
        \begin{overpic}[width=0.85\textwidth]{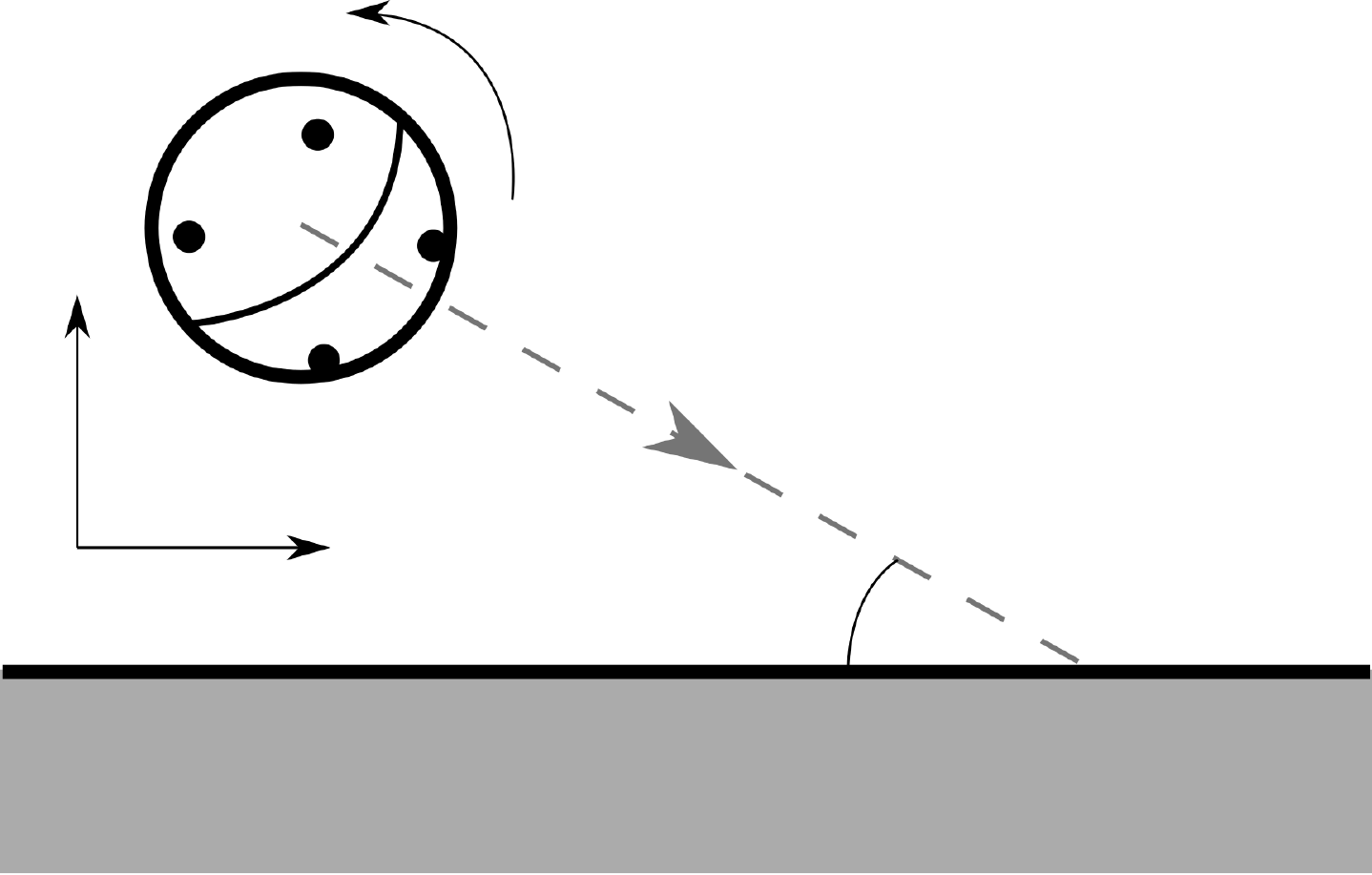}
		\put(38,58){$\omega$}
		\put(22,20){$x$}
		\put(2,40){$y$}
		\put(68,17){$\alpha$}
		\put(15,5.5){Bounce surface}
	\end{overpic}
    \caption{ }
    \label{fig:ic_setting}
	\end{subfigure}
	\caption{Left: A combination of frames recorded for a single
          bounce. Time intervals between frames vary and are chosen only for ease of illustration. Right: Definition sketch of the variables used throughout the analysis of the data.}
	
\end{figure}

On leaving the launcher, a high speed camera was triggered
manually or by a laser 
sensor to capture a video of the bounce. The recording was made using a 
Phantom VR603 high speed camera, recording at between 5,000 and 10,000 frames
per second. The camera was set to align the image
plane with the plane of motion of the ball. 
Each video clip consists of a number of frames prior
to and subsequent to the bounce. A superposition of frames from one such clip is shown  in Figure \ref{fig:ball_trace}.

For the purposes of automated tracking, the bounce was recorded against a plain 
black background. Furthermore, each ball was marked with a line along its seam and dots placed in a regular array around this seam. 
Videos were taken outdoors, in summer, for which natural lighting proved to be sufficient.

The settings on the launcher (wheel speeds and launch angle)
were chosen to span a wide range of conditions typical to golf.
Ball bounces for each setting
of the launcher were repeated 6 times in Campaign A and either 6 or
12 times in Campaign B. A total of 330 bounces
were recorded in Campaign A and 693 in Campaign B.

%

\begin{table}[!b]
	\centering
	\caption{Range of initial conditions for Campaigns A and B}
	\label{tab:ic_camA}
	\begin{tabular}{|lll|l|lll|}
	    \hline
        \multicolumn{7}{|c|}{\textbf{Campaign A}}  \\ \hline
        \multicolumn{7}{c}{ }  \\ \cline{1-3} \cline{5-7} 
		\multicolumn{3}{|c|}{\textbf{DIMENSIONAL QUANTITIES}} &  & 
		\multicolumn{3}{c|}{\textbf{SCALED QUANTITIES}} \\ \cline{1-3} 
		\cline{5-7} 
		\multicolumn{1}{|p{4cm}|}{} & \multicolumn{1}{p{1.2cm}|}{min} & 
		\multicolumn{1}{p{1.2cm}|}{max} &  & \multicolumn{1}{p{4cm}|}{} & 
		\multicolumn{1}{p{1.2cm}|}{min} & \multicolumn{1}{p{1.2cm}|}{max} \\ 
		\cline{1-3} 
		\cline{5-7} 
		\multicolumn{1}{|l|}{Speed {[}$\mathrm{m\,s}^{-1}${]}} & 
		\multicolumn{1}{l|}{6.35} & 56.1 &  & \multicolumn{1}{l|}{$\dot{x} \, 
		[\mathrm{s}^{-1}]$} & \multicolumn{1}{l|}{74.9} & 2260 \\ \cline{1-3} 
		\cline{5-7} 
		\multicolumn{1}{|l|}{Angle of incidence $\ga$ {[}deg{]}} & 
		\multicolumn{1}{l|}{18.3} & 76.4 &  & \multicolumn{1}{l|}{$\dot{y} \,  
		[\mathrm{s}^{-1}]$} & \multicolumn{1}{l|}{-1880} & -225 \\ \cline{1-3} 
		\cline{5-7} 
		\multicolumn{1}{|l|}{Spin {[}rot per min{]}} & 
		\multicolumn{1}{l|}{-3750} & 9100 &  & \multicolumn{1}{l|}{$\omega \, 
		[\mathrm{rad\,s}^{-1}]$} & \multicolumn{1}{l|}{-392} & 953 \\ 
		\cline{1-3} \cline{5-7} 
		 \multicolumn{7}{c}{ }  \\ 
		  \hline
        \multicolumn{7}{|c|}{\textbf{Campaign B}}  \\ \hline
        \multicolumn{7}{c}{ }  \\ \cline{1-3} \cline{5-7} 
        \multicolumn{3}{|c|}{\textbf{DIMENSIONAL QUANTITIES}} &  & 
		\multicolumn{3}{c|}{\textbf{SCALED QUANTITIES}} \\ \cline{1-3} 
		\cline{5-7} 
		\multicolumn{1}{|p{4cm}|}{} & \multicolumn{1}{p{1.2cm}|}{min} & 
		\multicolumn{1}{p{1.2cm}|}{max} &  & \multicolumn{1}{p{4cm}|}{} & 
		\multicolumn{1}{p{1.2cm}|}{min} & \multicolumn{1}{p{1.2cm}|}{max} \\  
		\cline{1-3} \cline{5-7} 
		\multicolumn{1}{|l|}{Speed {[}$\mathrm{m\,s}^{-1}${]}} & 
		\multicolumn{1}{l|}{1.93} & 38.7 &  & \multicolumn{1}{l|}{$\dot{x} \, 
		[\mathrm{s}^{-1}]$} & \multicolumn{1}{l|}{1.10} & 1720 \\ \cline{1-3} 
		\cline{5-7} 
		\multicolumn{1}{|l|}{Angle of incidence $\ga$ {[}deg{]}} & 
		\multicolumn{1}{l|}{16.4} & 89.5 &  & \multicolumn{1}{l|}{$\dot{y} \,  
		[\mathrm{s}^{-1}]$} & \multicolumn{1}{l|}{-1570} & -90.0 \\ \cline{1-3} 
		\cline{5-7} 
		\multicolumn{1}{|l|}{Spin {[}rot per min{]}} & 
		\multicolumn{1}{l|}{-3880} & 16200 &  & \multicolumn{1}{l|}{$\omega \, 
		[\mathrm{rad\,s}^{-1}]$} & \multicolumn{1}{l|}{-407} & 1670 \\ 
		\cline{1-3} \cline{5-7} 
	\end{tabular}
\end{table}

\subsection{Landing conditions}

The variables used to define the ball's landing conditions are given in
Figure \ref{fig:ic_setting}. Note
that $y$ is measured vertically upwards, meaning that landing 
velocities $\dot{y}$ will always be negative, and the spin
$\omega$ is 
measured anti-clockwise; thus $\omega>0$ represents backspin
and $\omega<0$ topspin. To provide
consistent and comparable measures of error, we use scaled
units that give approximately comparable values for linear and
rotational velocity;  thus, $x$ and $y$ are measured in units of
ball radii and  $\omega$ in radians per second. 

\begin{figure}
	\centering
	\begin{subfigure}{0.45\linewidth}
		\includegraphics[width = \linewidth]{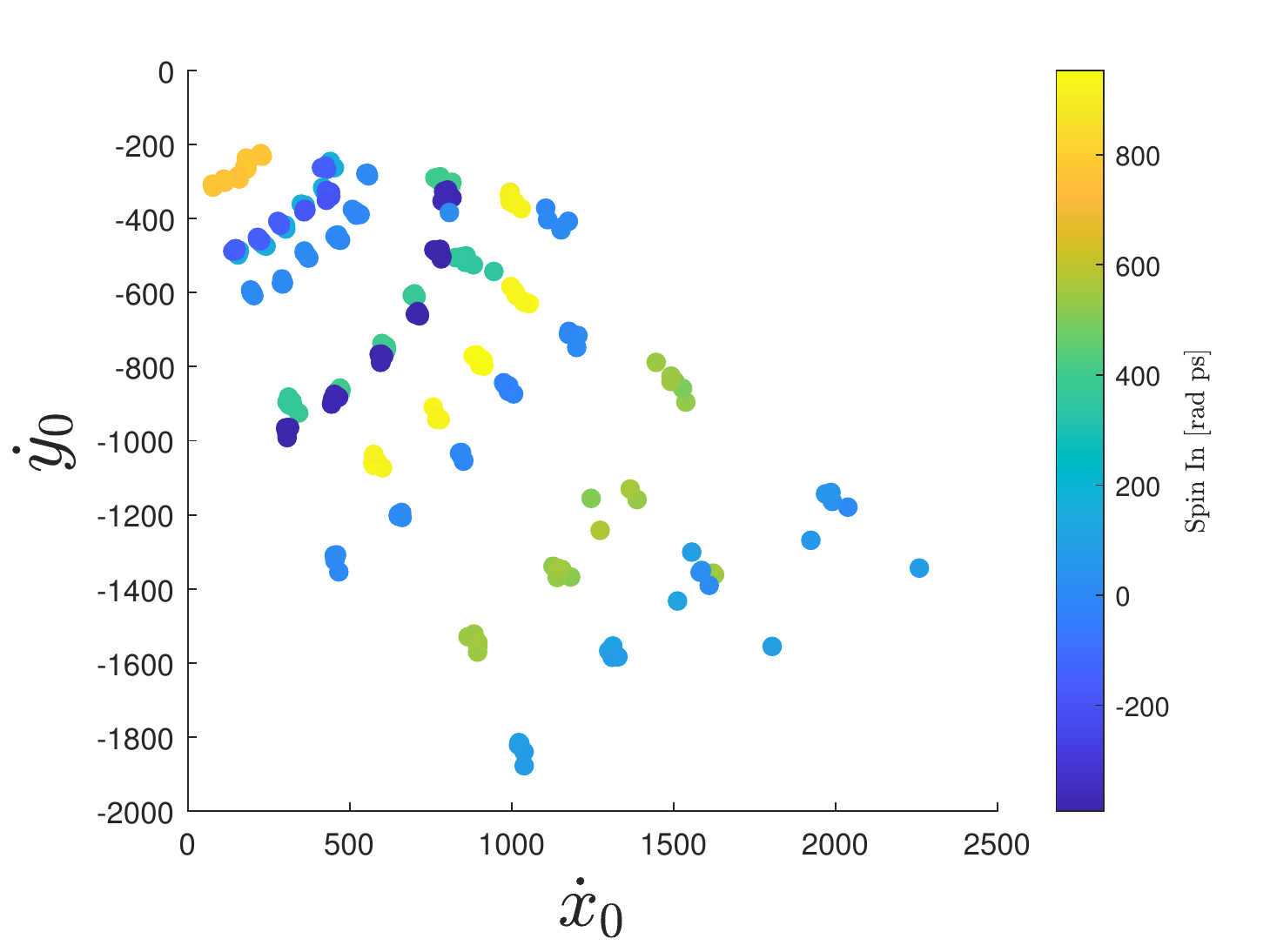}
		\caption{ }
		\label{sfig:campA_ic}
	\end{subfigure}
	\begin{subfigure}{0.45\linewidth}
		\includegraphics[width = \linewidth]{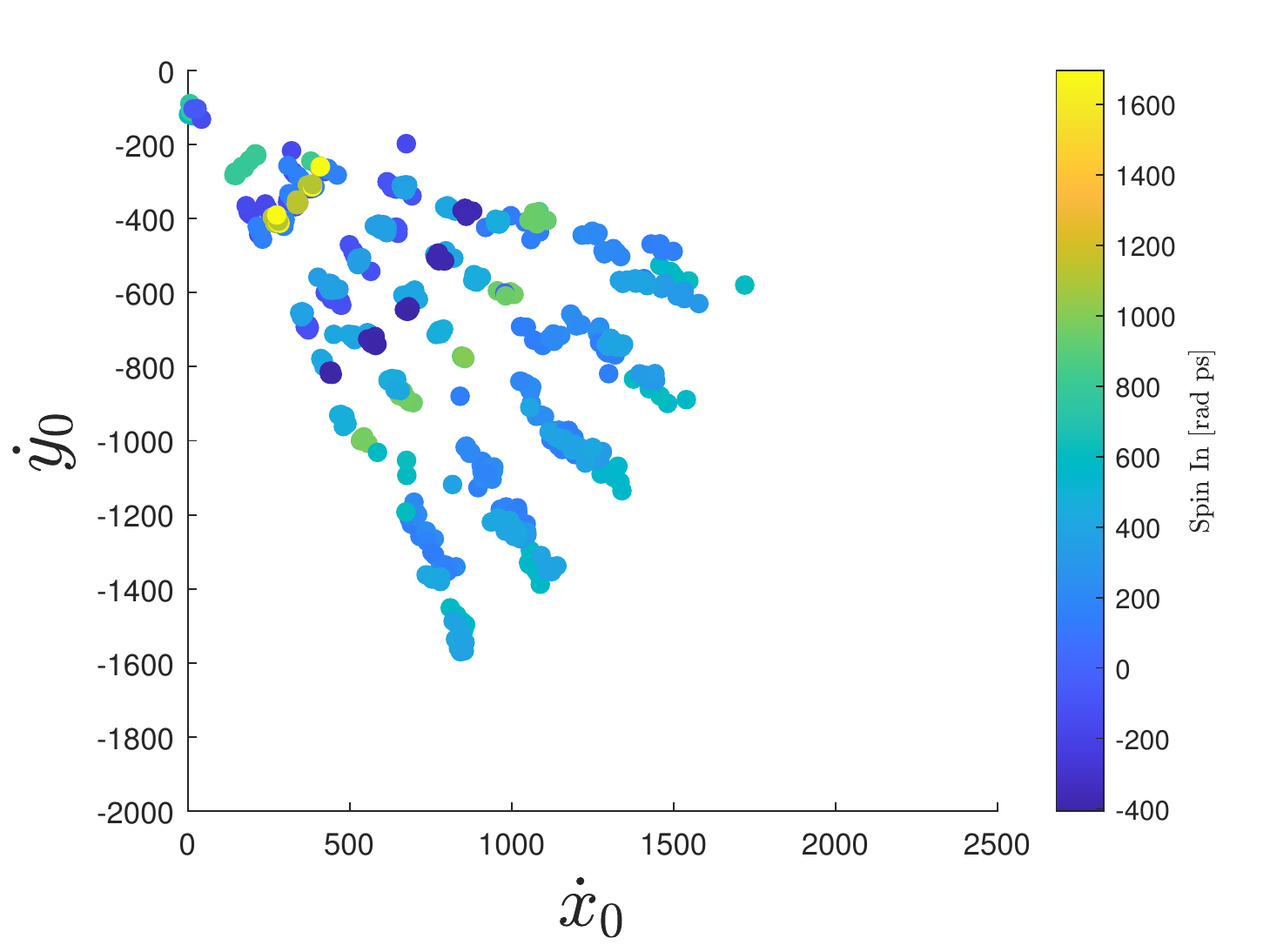}
		\caption{ }
		\label{sfig:campB_ic}
	\end{subfigure}
    \begin{subfigure}{0.45\linewidth}
		\centering
		\includegraphics[width = \linewidth]{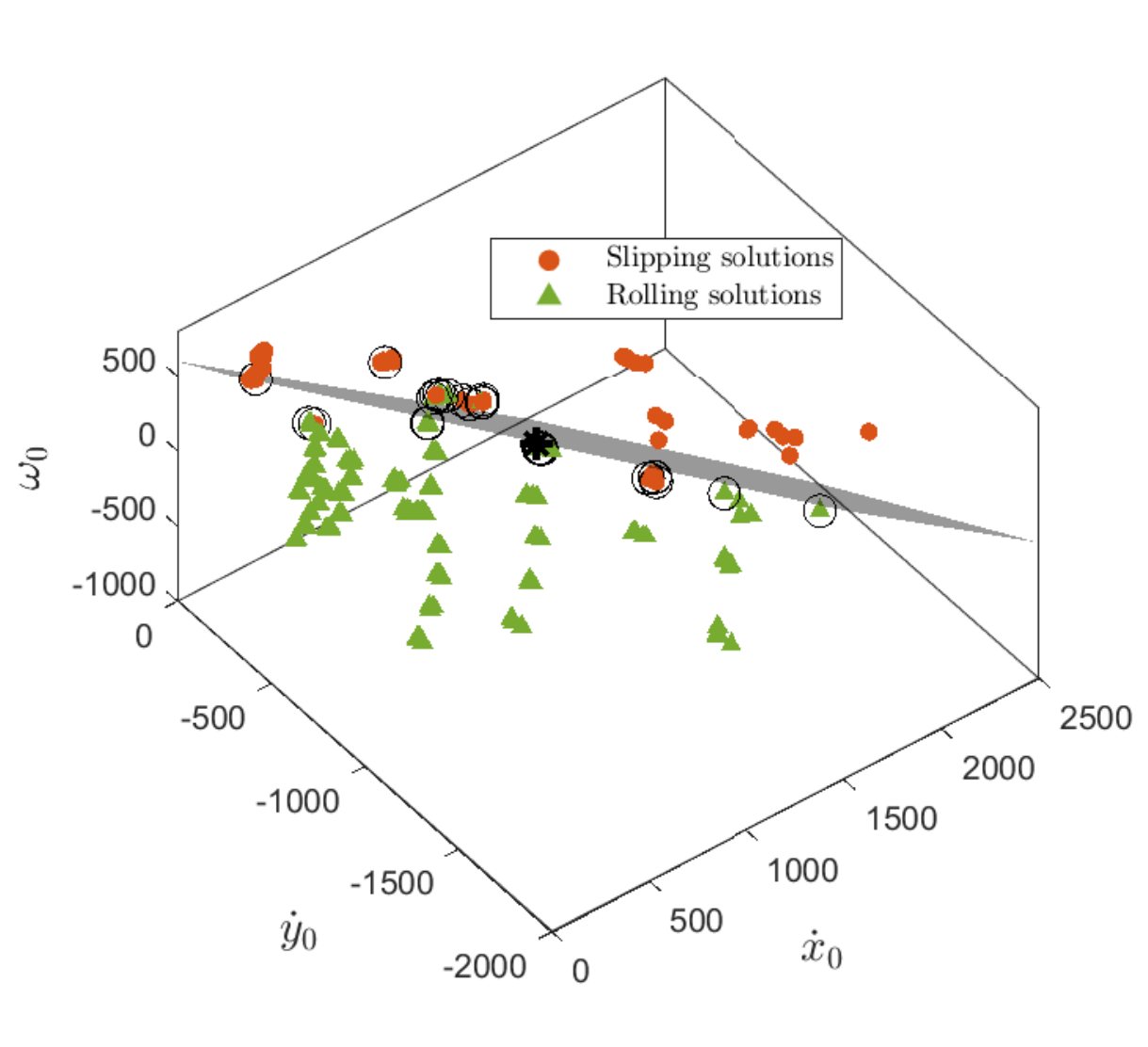}
		\caption{ }
	\end{subfigure} 
	\begin{subfigure}{0.45\linewidth}
		\centering
		\includegraphics[width = \linewidth]{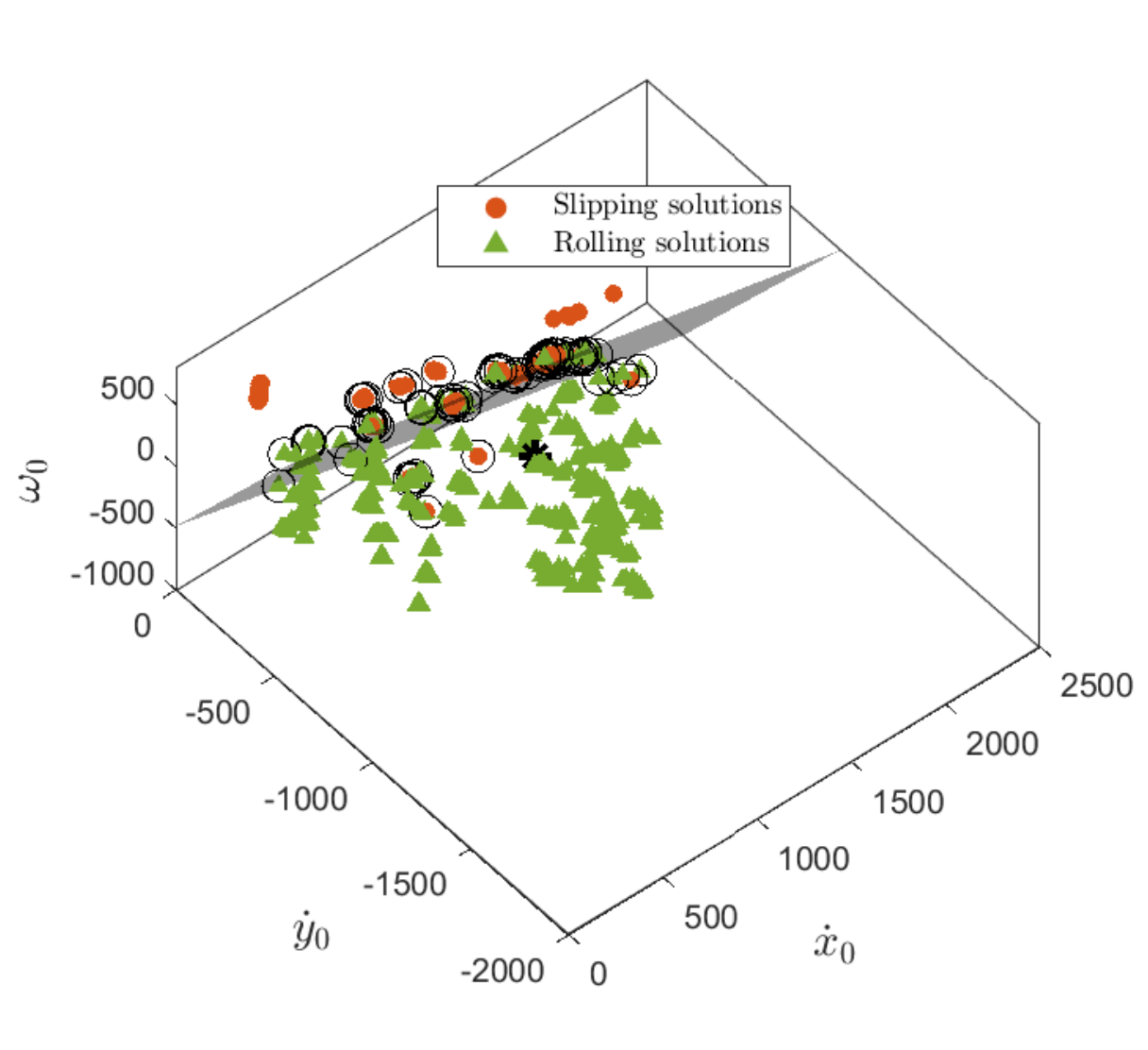}
		\caption{ }
	\end{subfigure}
	\caption{ An overview of the landing conditions for (a) 
	Campaign A and (b) 
	Campaign B. (c,d) Corresponding categorisation of the data based on the tangential velocity at lift off
	their initial conditions for (c) Campaign A; (d) Campaign B. Circled are 
	the boundary data points used for identifying the separating hyperplane.  The black asterisk marks typical landing conditions from a tee shot with a driver; speed $93.6$ feet per second, landing angle of $37.3^{\circ}$ and a backspin of $34.9$ revolutions per second.
 }
	\label{fig:ics}
\end{figure}

Table \ref{tab:ic_camA} presents the span of the landing conditions,
both in unscaled and scaled co-ordinates.  The full range of
the initial condition space can be seen in Figure \ref{fig:ics} for each campaign. The data appears clustered as a result of discrete dials used to control the launcher's settings. 

\subsection{Data extraction} \label{sec:algorithm}

A bespoke 
procedure was developed in
\textsc{Matlab} \cite{MATLAB:2010}
for the 
extraction of data from the videos. First, the ball was located in each frame (and frames where 
the ball was not found were removed from consideration), using the 
circle detection function 
\texttt{imfindcircle} which is based on the Hough transform \cite{Hough_transform}. From the change in co-ordinates of the centre between frames, the $x$ and $y$ velocities were established.
Second, for each frame, a reduced image was  was taken, centred
around the ball, with a small margin around it. We then used
\textsc{Matlabs}'s 
\texttt{extractFeatures} and \texttt{matchFeatures} functions to respectively 
identify significant features to match them between frames. 
This, in turn, allowed us to construct a 2D rotation matrix to estimate the degree of rotation between frames. Although automated, it was typically found that the black dots on the ball were the mostly commonly extracted and matched features. 

\begin{figure}
	\centering
	\begin{subfigure}{0.45\linewidth}
		\centering
		\begin{overpic}[width=\textwidth]{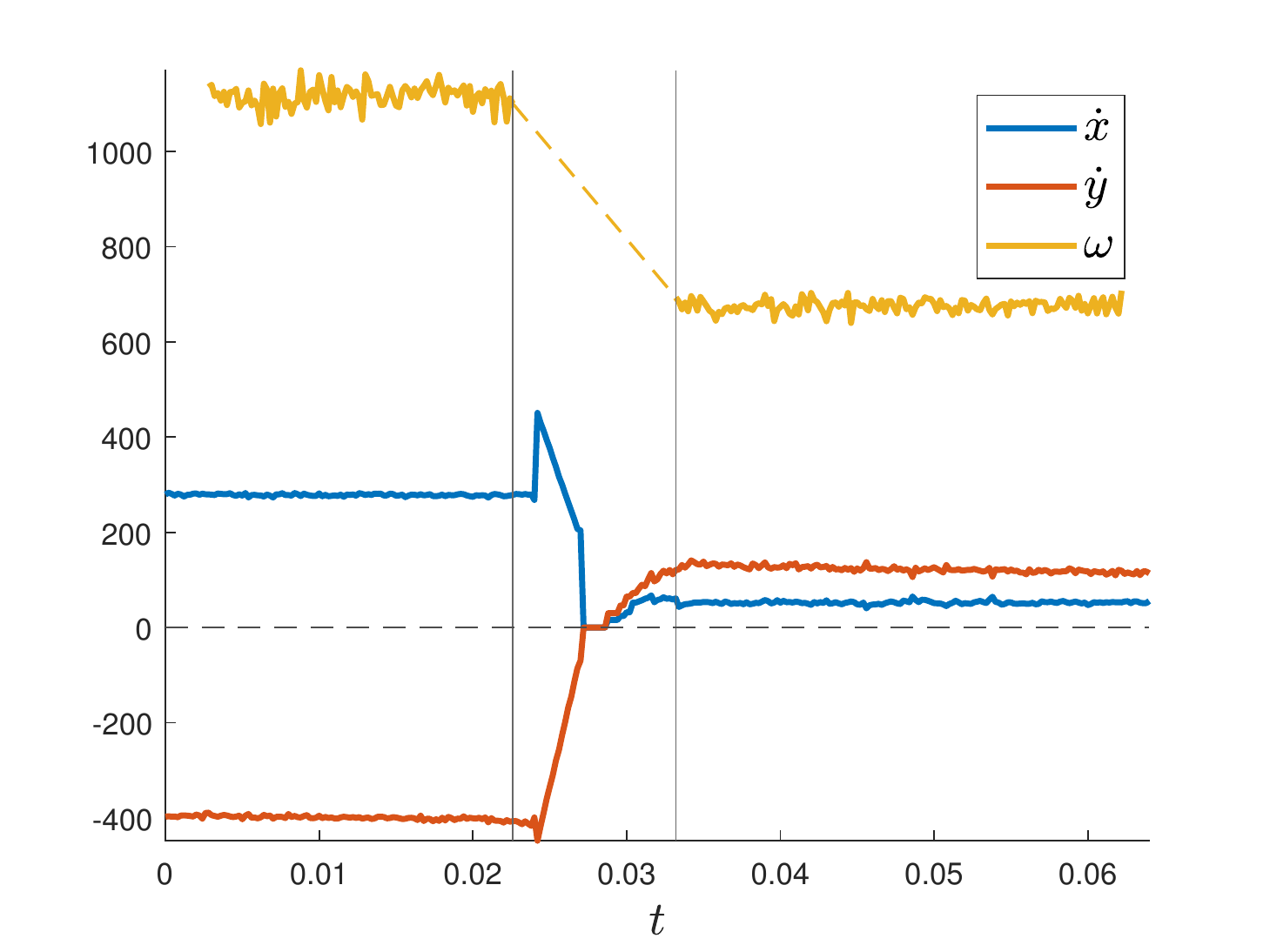}
			\put(41,45){\tiny Bounce}
		\end{overpic}
		\caption{ }
	\end{subfigure}
	\begin{subfigure}{0.45\linewidth}
		\centering
		\begin{overpic}[width=\textwidth]{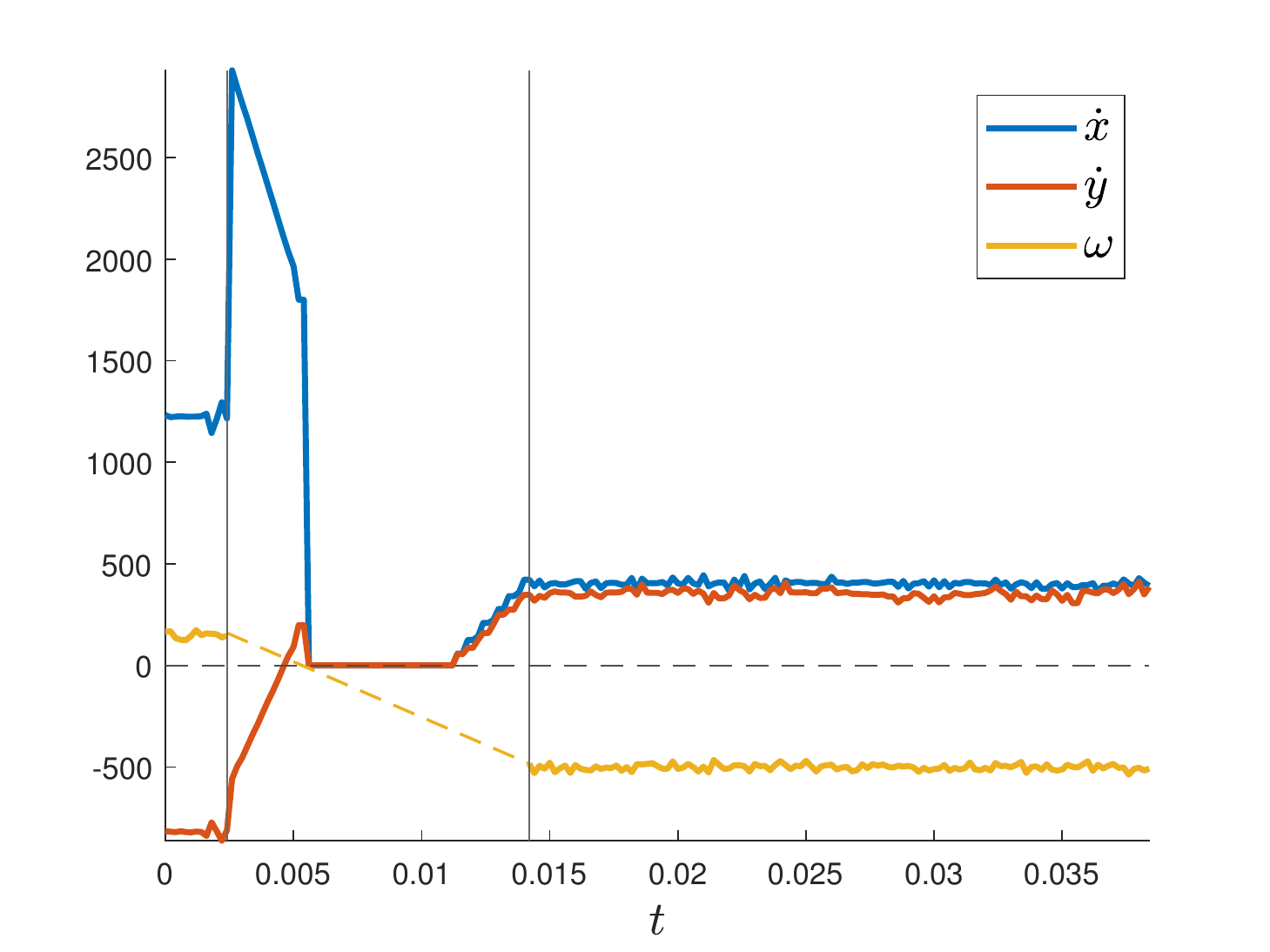}
			\put(25,45){\tiny Bounce}
		\end{overpic}
		\caption{ }
	\end{subfigure}
	\caption{Unprocessed tracing data of a ball prior, during and after bounce 
	for two different launch conditions.}
	\label{fig:raw_tracking}
\end{figure}

In all cases, contact of the ball with the surface occurred over a duration of many frames. The initiation and end time of a
bounce was estimated from the  
identification of abrupt changes in velocity of the lowest visible 
point of the ball. Also, during bounce, the view of 
the ball is obstructed by the turf, and locating the ball in the frame and 
estimating its spin became unreliable (see e.g.~Figure \ref{fig:raw_tracking}).
For that reason, we decided to restrict the ball bounce
analysis to inbound and outbound conditions, just prior and subsequent to the
bounce. 

\subsection{Data analysis}

Each bounce was categorised using six quantities; 
the horizontal velocity, vertical velocity 
and spin for both touch down  (denoted as $\dot{x}_0, \dot{y}_0$ and 
$\omega_0$ respectively) and lift-off ($\dot{x}_F, \dot{y}_F$ and 
$\omega_F$ respectively). 

\subsubsection{Measurement error}

As indicated in Fig.~\ref{fig:raw_tracking}, the data extracted 
was not a single point measurement of each
velocity and spin, but a noisy time series of measurements for
each variable; evaluated at times $t_1, t_2, \dots, t_N$, where the time interval 
$\Delta t = t_{i+1}-t_i$ is the frame rate, and is
constant for all $i = 1, \dots N-1$. For the ball velocities, we used the line of best fit to the 
derivatives of the position measurements. That is, for the horizontal velocity we found
\begin{equation}
v_x(t) = m\,t +c 
\end{equation}
such that
\begin{equation}
	\ve_x = \sum_{i=1}^{N-1} \left|\frac{x_{i+1} - x_{i}}{\Delta t } - 
	v_x(t_{i+1}) \right|^2
\end{equation}
is minimised. The vertical velocity estimate, $v_y(t)$,
was calculated similarly. 
Incorporating a linear fit accounted for possible effects of the 
acceleration, such as gravity in the case of the vertical velocity or drag in the 
case of the horizontal velocity. Parameters $m$ and $c$ were estimated together 
with their 95\% confidence intervals (CI), which then yielded the 
95\% prediction intervals (PI) for the values of $v_x$ and $v_y$ evaluated at 
the time of impact or lift-off. The size of the PI gives n parsimonious estimate of measurement error for each velocity in each trial. 

It was found useful to treat spin differently, as no evidence of linear
changes in spin was found during the recorded short time intervals before and after bounce. We thus assumed that recorded spin in each interval  follows 
a normal distribution, with the noise attributed to measurement error. A normal distribution was thus fit to time series for each spin measurement and the mean 
of this distribution taken as the measured spin.
The 95\% CIs of the mean is thus taken as the
PI for spin. 

\subsubsection{Repeatability}

Alongside estimates of measurement error provided by the PI for each launch,
using 
6 or 12 repetitions for each launching setting enabled a check on whether the 
tests recorded with similar initial conditions led to similar outbound measurements.
For each test condition, the mean and 95\% CI's on the mean were calculated for each 
quantity (horizontal velocity $\dot{x}$, vertical velocity $\dot{y}$ and spin $\omega$),
both on their inbound and on outbound measurements. A larger width of each CI
shows a larger spread of data around the mean and less repeatability in that quantity.

\subsubsection{Distinguishing between slipping and rolling}
\label{sec:slipping_rolling}

The tangential velocity of the lowest point of the golf ball should carry 
information about the frictional dynamics of the ball. If the 
velocity of this point is zero, the frictional interface will be in a state of `stick'
which suggests that the ball is rolling for a time. In contrast, a non-zero tangential velocity suggests that the  ball is `slipping' against the surface. 
Denoting such a point as $P$, in the scaled variables, 
the tangential velocity of $P$ is given by
	\begin{equation}
		v_P = \dot{x} + \omega.
	\end{equation}
	
	
We separated the initial condition space into two groups,
based on the measurement of the tangential velocity of $P$ at touch down and at lift off. If the tangential velocity were of the same sign at touch down and lift-off (non-zero in each case) then it was deemed that the ball was
{\bf slipping} throughout the bounce. In all other cases, we deduced that
the ball must have entered {\bf rolling} at some
point during the bounce.

To identify a possible separation manifold between these groups, a support vector machine (SVM) 
classifier was trained using the sequential minimal optimisation (SMO) algorithm \cite{smo}. 
The trained SVM identified the data points at the boundary between the two classes, and these 
were used to identify the surface separating the landing condition space into the subspaces 
generating the rolling and slipping trajectories. 
On experimentation (and using insights obtained with models in \cite{biber_analysis}), we found that the boundary was well approximated by a plane within
landing condition space, for the range considered.

\subsection{Comparison with existing models}

A key aim of this paper is to compare the data with some of the
most commonly used models in the literature. 

\subsubsection{Rigid-bounce model}

The simplest model (e.g.~\cite{Daish}), assumes that the bounce
is instantaneous and results in a proporational loss of normal velocity,
with the ratio
\begin{equation}
r = \left| \dot{y}_0/\dot{y}_F \right|,
\end{equation}
known as Poisson's coefficient of restitution.
Further, Coulomb 
friction is assumed to act between the ball and the surface, with a 
constant coefficient $\mu$, such that the ball is
$$
\mbox{slipping when } \left|\lambda_T \right| = \mu \lambda_N \mbox{ and rolling when } 
\left|\lambda_T \right| < \mu \lambda_N,
$$
where $\lambda_T$ and $\lambda_N$ denote the respective
tangential and normal forces acting on the point $P$. The model thus consists 
of two parameters, $\mu$ and $r$, which are properties of the bounce surface.
Typically $0<r <1$, $0<\mu<1.5$

Depending on whether the ball enters rolling or slips throughout the impact, the lift-off values predicted by the model are easily found to be 
\begin{equation}\label{eq:rigid_bounce}
	\begin{matrix}
		\dot{x}_F = \dot{x}_0 + \mu(1+r) \dot{y}_0,\, & \dot{y}_F = 
		-r\dot{y}_0\,\ & \omega_F = \omega_0 + \frac{5}{2}\mu(1+r) \dot{y}_0& 
		\quad\mbox{if } -\frac{\dot{x}_0+ \omega_0}{\dot{y}_0}> \frac{7}{2} \mu 
		(1+r);\\
		\dot{x}_F = \frac{5}{7} \dot{x}_0 - \frac{2}{7} \omega_0,\, & \dot{y}_F 
		= -r\dot{y}_0\,\ & \omega_F = -\frac{5}{7} \dot{x}_0 + \frac{2}{7} 
		\omega_0&  \quad\mbox{if } \left|\frac{\dot{x}_0+ 
		\omega_0}{\dot{y}_0}\right| < \frac{7}{2} \mu (1+r);\\
		\dot{x}_F = \dot{x}_0 - \mu(1+r) \dot{y}_0,\, & \dot{y}_F = 
	-r\dot{y}_0\,\ & \omega_F = \omega_0 - \frac{5}{2}\mu(1+r) \dot{y}_0& 
	\quad\mbox{if } -\frac{\dot{x}_0+ \omega_0}{\dot{y}_0} < - \frac{7}{2} \mu 
	(1+r).
	\end{matrix}
\end{equation}

The parameters $\mu$ and $r$ were fit from the available data using a
structured approach. First, the coefficient of restitution was found
using a least-squares approach applied to each measured inbound and
outbound vertical velocity.  Second, a least squares
estimate was used again to find $\mu$.
Equations \eqref{eq:rigid_bounce}
were then used to calculate predicted values
\begin{equation}
  \vec{P}_F^i = [\dot{X}^i, \dot{Y}^i,
  \Omega^i]^{\intercal}  
\end{equation}
for $[\dot{x}_F, \dot{y}_F,
  \omega_F]^{\intercal}$ for
each measured landing sample $i$ of incoming values $\vec{p}_0^i =
[\dot{x}_0^i, \dot{y}_0^i, \omega_0^i]^{\intercal}$.

The error of the model for all $N$ samples in a campaign was calculated as:
\begin{equation}\label{eq:error}
	\delta = \frac{1}{N}\sum_{i=1}^{N} \frac{\lVert \vec{p}_F^i - 
	\vec{P}_F^i	\rVert}{\lVert \vec{p}_0^i\rVert},
\end{equation}
where 
$\vec{p}_F^i = [\dot{x}_F^i, \dot{y}_F^i, \omega_F^i]^{\intercal}$ 
are the measured lift-off values 
and 
$\lVert \cdot \rVert$ represents the usual Euclidean norm of a vector. 
Note that to be a meaningful error measurement, each component of  $\vec{p}_0, \vec{p}_F, \vec{P}_F$ must be of the same order 
of magnitude, highlighting the importance of working with our scaled
velocity quantities.  The error measure computed in \eqref{eq:error} can be thought
of as the root-mean-square error with respect to the
initial kinetic energy of the system. 

Similarly, we computed the average error for each individual quantity $\dot{x}, \dot{y}, \omega$, (generically denoted as $q$ with a predicted computed value $Q$) via:
\begin{equation}\label{eq:error_individual}
	\Delta q = \frac{1}{N}\sum_{i=1}^{N} \frac{\lvert q_F^i - Q_F^i 
		\rvert}{\lVert \vec{p}_0^i\rVert}.
\end{equation}  
Note that the average quantities computed by \eqref{eq:error_individual}
neither sum nor average to the error computed by \eqref{eq:error} --
they are simply an indicator of  where the largest error is generated.

A fixed estimate of the coefficient of restitution $r$ for the rigid
bounce model was calculated by varying parameter $\mu$, such
that the total error, as specified by Equation \eqref{eq:error}, is
minimised.

\subsubsection{Inclined surface model}

Penner's model \cite{penner2002} was examined in a similar fashion. This model proposes that 
the bounce of a rigid ball against a horizontal compliant surface can be modelled as the bounce of a rigid ball against a rigid surface that has been rotated by an angle $\beta$.
The specific formulation proposed in \cite{penner2002} is that
\begin{equation}\label{eq:penner_angle}
	\beta =k_P s\, \phi, \quad 
 \mbox{where } s=\sqrt{\dot{x}_0^2+\dot{y}_0^2} \quad \mbox{and  } \phi = \mbox{arctan}\left[-\frac{\dot{y}_0}{\dot{x}_0}\right]. 
 \end{equation}
Here $s$ is the incoming ball speed and $\phi$ is its landing angle; 
$k_P$ is a turf-dependent constant.   

The model requires fitting parameter $k_p$, in addition to $\mu$ and $r$.  
We ran Matlab's genetic algorithm toolbox
\texttt{ga} to fit all three parameters, minimising the error function \eqref{eq:error}. The chosen tolerance in 
minimising the function was $10^{-15}$. In each setting, the
coefficient of restitution was specified within the frame of
reference that is rotated through $\beta$. 

We explored the idea of the rotated frame of reference in
two ways -- either the angle of rotation was determined by the
initial condition, as specified by \eqref{eq:penner_angle},
or supposing that the angle $\beta$ is fixed for all initial conditions.

\subsubsection{A piecewise-linear data fit}

In addition, within each half-space, either side of the boundary between slipping and rolling
identified in Section \ref{sec:slipping_rolling},
we attempted to fit an affine transformation. That is, for the  
measured initial condition $\vec{p}_0$ and outbound $\vec{p}_F$, we approximate the latter with
\begin{equation}\label{eq:piecewise_fit}
\vec{P}_F = A \vec{p}_0 + \vec{b},
\end{equation}
where $A$ is a $3\times3$ matrix, and $\vec{b}$ is a 3-dimensional vector.

To allow for validation of the fit, we separated our data into
two categories for each campaign. 
One data point from each launcher setting (chosen at random)
was removed from the general set of data and 
placed into a testing set, with the remaining data points used for training of the model \eqref{eq:piecewise_fit}. 
The testing data set was used to evaluate the accuracy of model \eqref{eq:piecewise_fit},
using the error measure \eqref{eq:error} and \eqref{eq:error_individual}.

\section{Results}

The full data set is presented in the Appendix \ref{sec:data_scatter} and is available online \cite{videos}
along with the full videos. The data set covers a wide range of
inbound velocity and spins, and includes examples of backwards bounce, that is where the sign of the horizontal velocity $\dot{x}$ reverses between impact and lift-off.

\subsection{Measurement error and repeatability}

\begin{figure}[!htb]
	\centering
	\begin{subfigure}{0.9\linewidth}
		\centering
		\includegraphics[width=\linewidth]{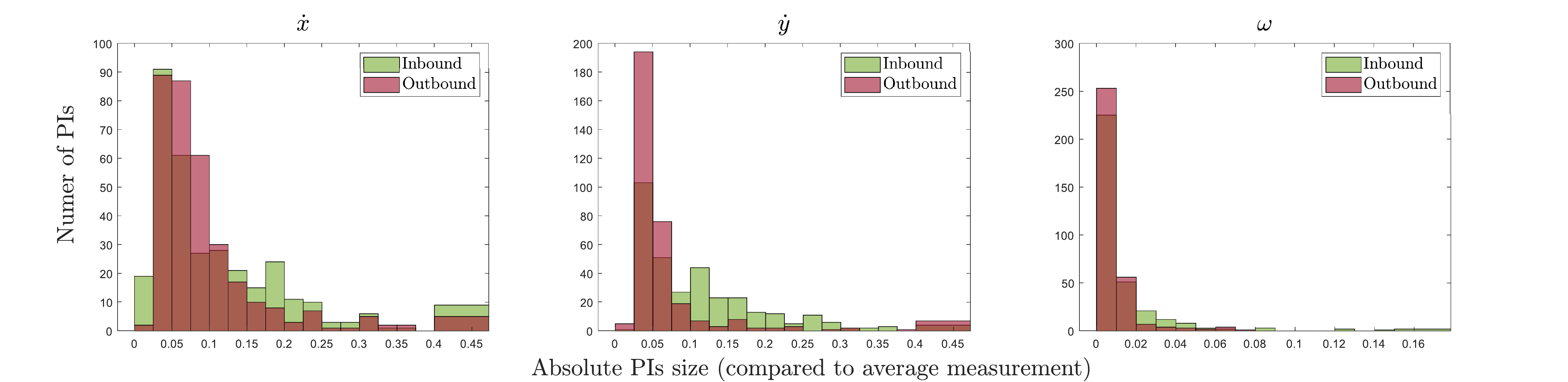}
		\caption{ }
	\end{subfigure}
	\begin{subfigure}{0.9\linewidth}
		\centering
		\includegraphics[width=\linewidth]{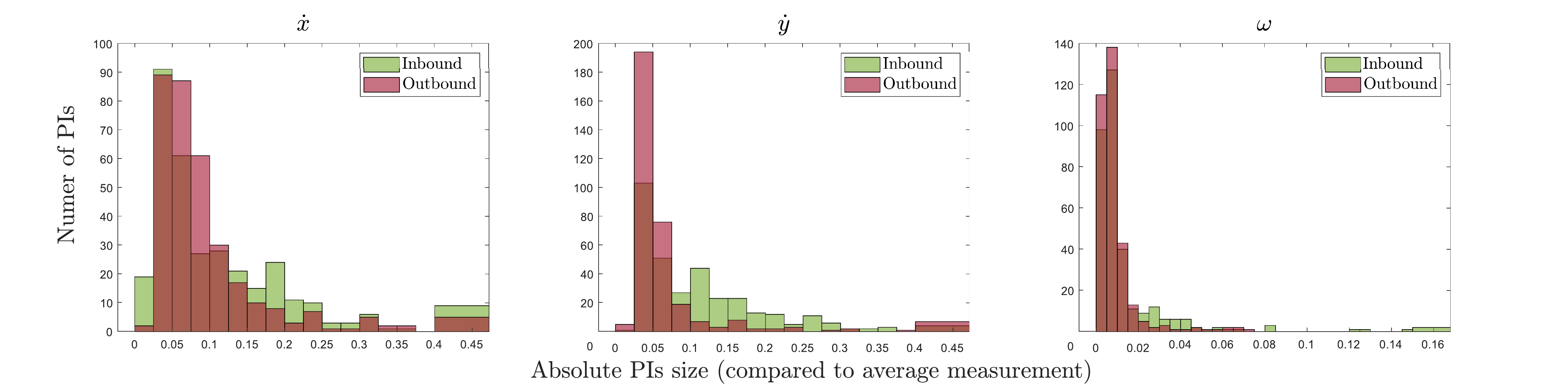}
		\caption{ }
	\end{subfigure}
	\caption{Histogram of widths of PIs in (a) Campaign A; (b) Campaign B. 
	values are rescaled by the 
	average measurement across all samples. From left to right these are for 
	the horizontal velocity $\dot{x}$, vertical velocity $\dot{y}$ and spin 
	$\omega$. Green bins represent measurements taken on the inbound, red bins 
	represent those taken on the outbound. }
	\label{fig:ci_hist}
\end{figure}

Figure \ref{fig:ci_hist} shows
histograms of the widths of the prediction intervals (PIs).
These are re-scaled with respect to the average measurement across the
samples. That is, a value $1$ on the horizontal axis of each histogram would correspond to the average horizontal, vertical or angular velocity on either inbound or outbound across all samples. 
These PI's provide a parsimonious estimate of the error in measuring speeds
via the video analysis. 

%

In contrast, the repeatability of tests with identical launcher settings
is illustrated in Figure
\ref{fig:repeatibility}.  Each circle denotes the mean of the
respective quantity for a particular test condition, where error bars
represent the 95\% confidence interval about the mean.


\subsection{Piecewise linear model slipping and rolling} \label{sec:tangential_vel}

\begin{figure}[!htb]
	\centering
	\begin{subfigure}{0.45\linewidth}
		\centering
		\includegraphics[width=\linewidth]{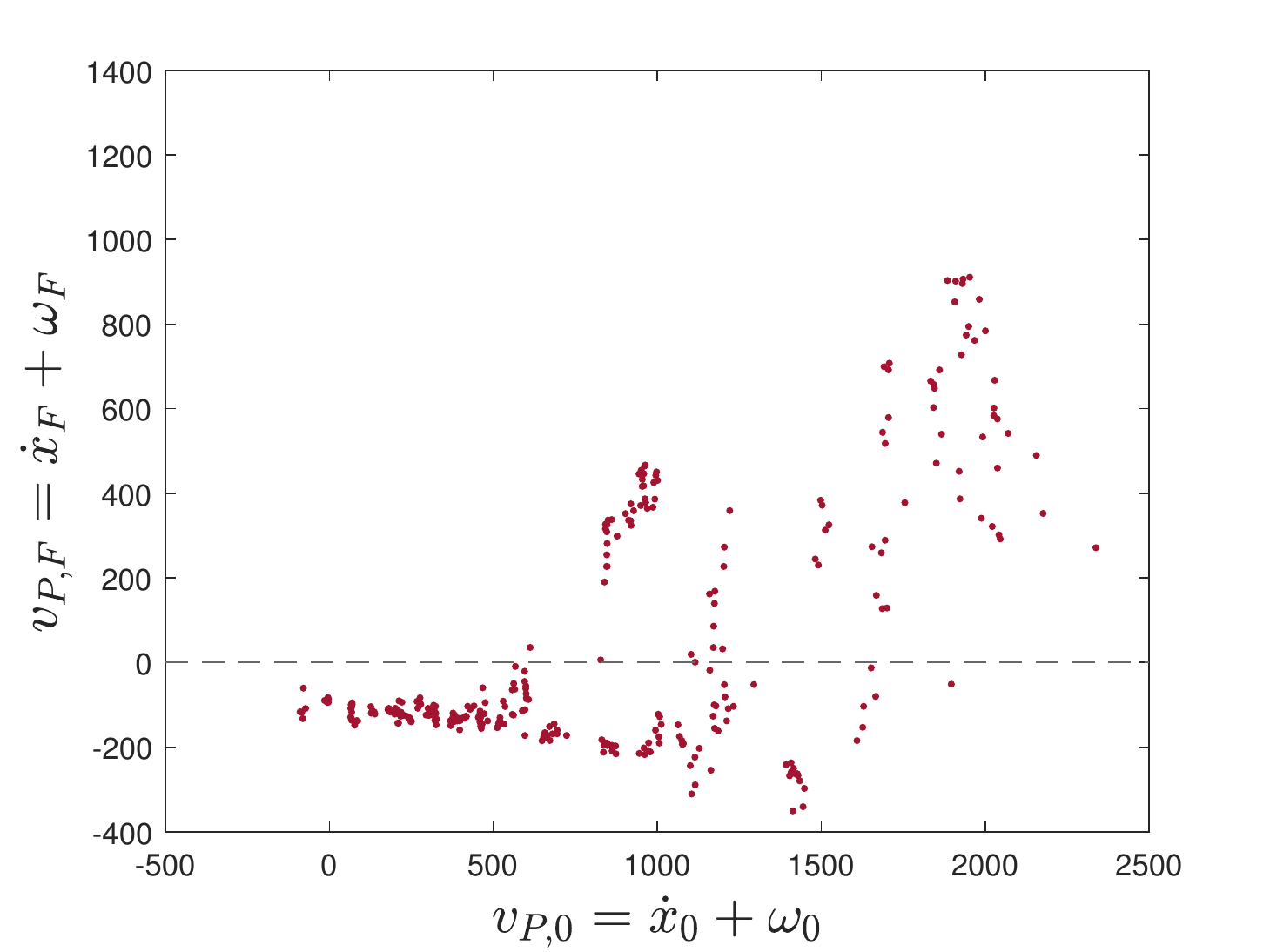}
		\caption{ }
	\end{subfigure}
	\begin{subfigure}{0.45\linewidth}
		\centering
		\includegraphics[width=\linewidth]{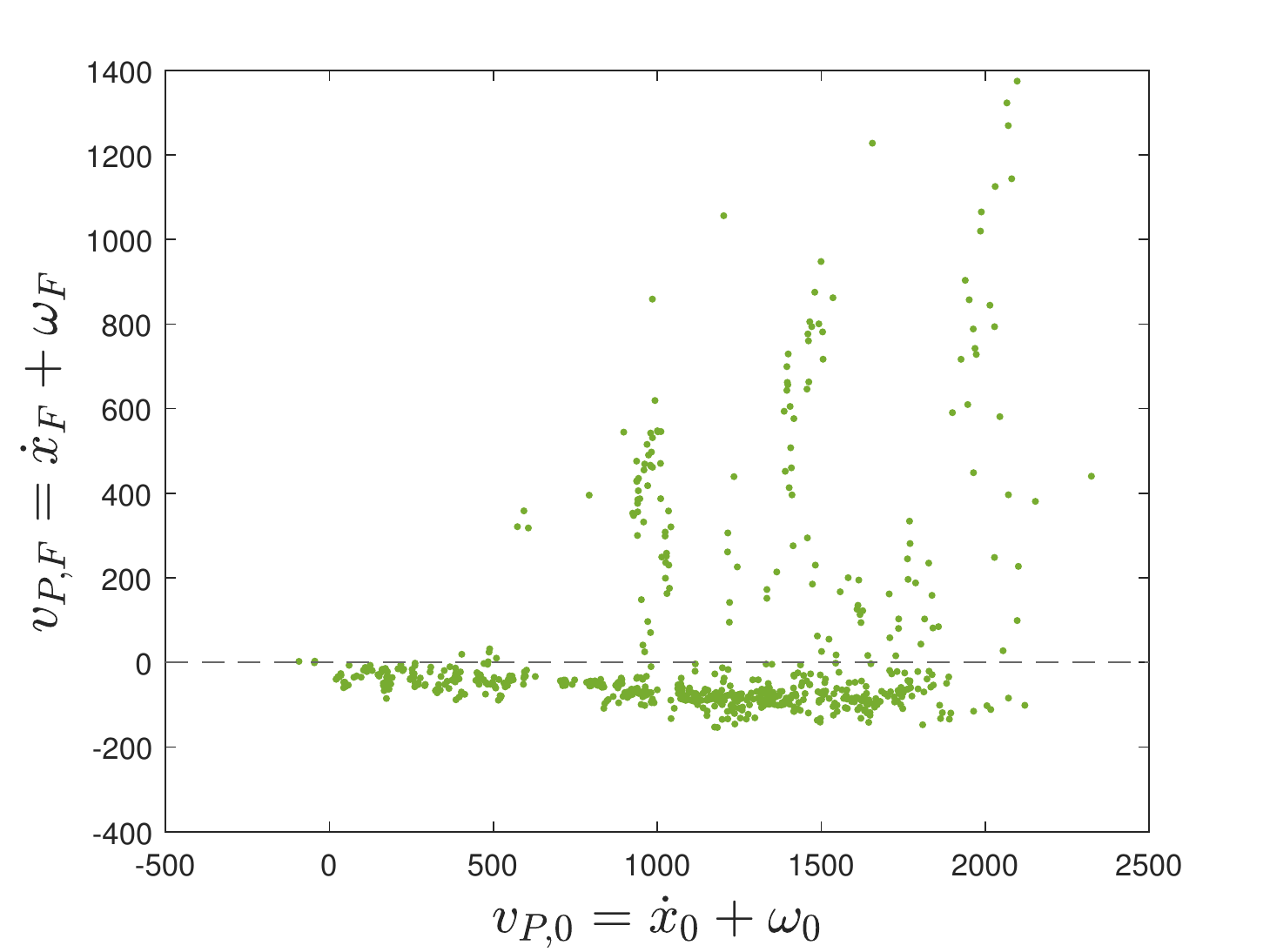}
		\caption{ }
	\end{subfigure}
	\caption{Tangential velocities at landing and lift off. Values for (a) 
	Campaign A; (b) Campaign B. Noted with the dashed line is the zero lift-off tangential velocity.}
	\label{fig:tang_vel}
\end{figure}

Figure \ref{fig:tang_vel} presents the tangential velocity $v_p$ of point $P$ at the time of
impact and  lift off for both campaigns. In particular, observe an
offset of lift-off velocities away from zero. That is, balls seem rarely, if ever, to lift off in a state of rolling.  There is always some slip at
lift off. 
Furthermore, notice a clear separation between those samples that lift off 
with positive tangential velocity $v_p$ from those with negative $v_p$.

The cases with $v_P>0$ correspond to cases with high initial backspin.
In such cases, the ball  slips throughout the impact and lifts off with 
backspin again. Conversely, the cases with $v_p<0$ at lift-off tend to
occur when either the ball landed with high
topspin ($v_p<0$ both at impact and lift-off) or where at some point 
during the bounce phase the ball entered rolling ($v_p<0$ initially, but $v_p>0$
at lift off). It is only the final of these cases that we refer to as a
`rolling' trajectory.

The distinction between rolling and slipping bounces is visualised in 
three-dimensional space in Figure \ref{fig:ics}. For each campaign, the 
data are separated into
different regions of the landing condition space, with only a few data points
being mis-labelled by the approximate dividing plane. The boundary cases are 
identified using the SVM algorithm, and a plane is fitted through the 
identified boundary points using least squares approximation.
Following separation of landing conditions
conditions with such a plane, an affine transformation between incoming and 
outgoing speeds can be identified for each of the subsets.


The resulting fits for Campaign A were:
\begin{equation}\label{eq:disjointA}
	\begin{bmatrix}
		\dot{x}_F \\ \dot{y}_F \\ \omega_F 
	\end{bmatrix} = \begin{cases}
	\begin{bmatrix}
		0.536 & -0.004 & -0.205 \\ 0.144 & -0.407 & -0.027 \\ -0.524 & 0.091 & 
		0.318
	\end{bmatrix} \begin{bmatrix} \dot{x}_0 \\ \dot{y}_0 \\ \omega_0 
	\end{bmatrix}+ \begin{bmatrix} -111 \\ 23.7 \\39.4 \end{bmatrix}  & 
	\mbox{if }  0.891 \dot{x}_0 +0.763 \dot{y}_0 +\omega_0 < 600\\
		\begin{bmatrix}
			0.674 & 0.107 & -0.005 \\ 0.103 & -0.390 & 0.017 \\ -0.153 & 
			0.277 & 1.01
		\end{bmatrix} \begin{bmatrix} \dot{x}_0 \\ \dot{y}_0 \\ \omega_0 
		\end{bmatrix}+ \begin{bmatrix} -154 \\ 25.0 \\ -267 \end{bmatrix}  & 
		\mbox{if }  0.891 \dot{x}_0 +0.763 \dot{y}_0 +\omega_0 > 600;
\end{cases}
\end{equation}
and similarly, for Campaign B the fits were:
\begin{equation}\label{eq:disjointB}
	\begin{bmatrix}
		\dot{x}_F \\ \dot{y}_F \\ \omega_F 
	\end{bmatrix} = \begin{cases}
		\begin{bmatrix}
			0.326 & 0.222 & -0.213 \\ 0.185 & -0.008 & -0.076 \\ -0.333 & 
			-0.147 & 0.224
		\end{bmatrix} \begin{bmatrix} \dot{x}_0 \\ \dot{y}_0 \\ \omega_0 
		\end{bmatrix}+ \begin{bmatrix} 175 \\ 90.4 \\ -171 \end{bmatrix}  & 
		\mbox{if }  0.377 \dot{x}_0 +2.25 \dot{y}_0 +\omega_0 < -485\\
		\begin{bmatrix}
			0.606 & 0.649 & -0.007 \\ 0.146 & -0.089 & -0.006 \\ -0.192 & 
			1.01 & 0.887
		\end{bmatrix} \begin{bmatrix} \dot{x}_0 \\ \dot{y}_0 \\ \omega_0 
		\end{bmatrix}+ \begin{bmatrix} 117 \\ 49.3 \\ 35.0 \end{bmatrix}  & 
		\mbox{if }   0.377 \dot{x}_0 +2.25 \dot{y}_0 +\omega_0 > -485.
	\end{cases}
\end{equation}

\begin{figure}
	\centering
	\begin{subfigure}{0.9\linewidth}
		\centering
		\includegraphics[width = \linewidth]{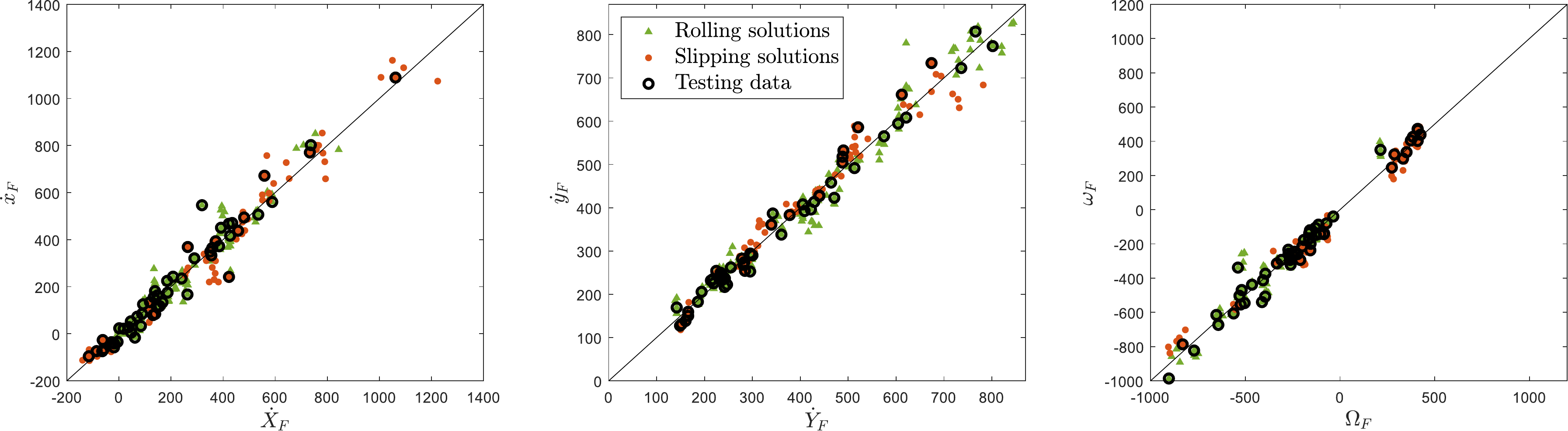}
		\caption{ }
	\end{subfigure}
	
	\begin{subfigure}{0.9\linewidth}
		\centering
		\includegraphics[width = \linewidth]{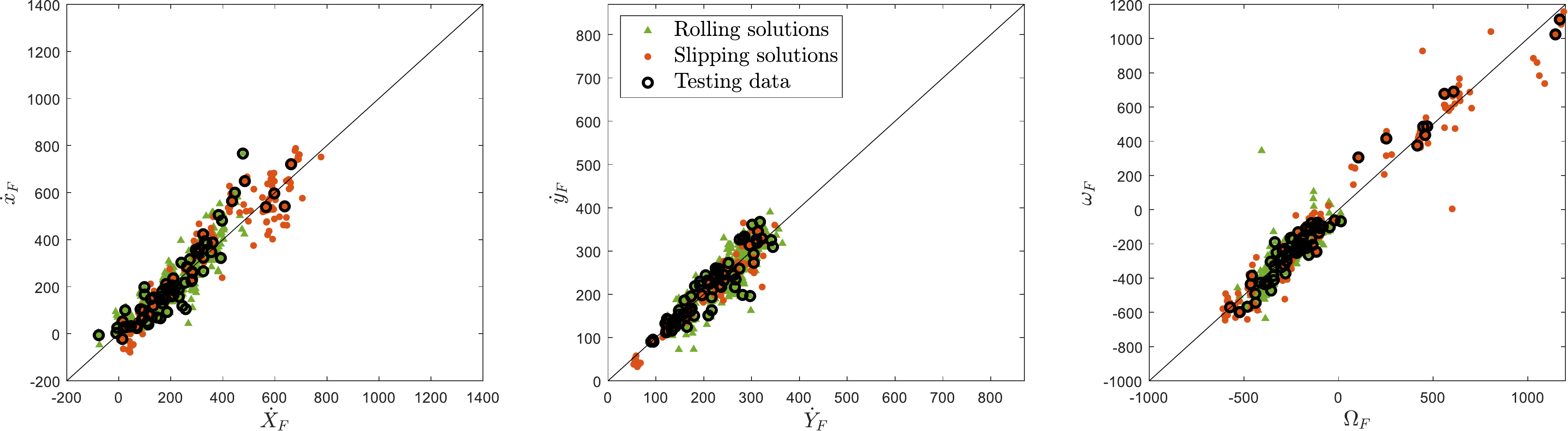}
		\caption{ }
	\end{subfigure}
	\caption{Comparison of the predicted lift off values against the observed 
	results. Solid line indicates the exact fit between prediction and 
	observation. Triangle markers indicate samples classified as rolling, and 
	the circular markers indicate slipping bounces. Marked with black circles 
	are the data samples randomly selected for the testing set to examine the 
	goodness of fits. Data is obtained from (a) Campaign A; (b) Campaign B.  }
	\label{fig:fit_goodness}
\end{figure}

The support vectors, which identify the boundary data, are highlighted in 
Figure \ref{fig:ics}. The error values for the presented fits are given in 
Table \ref{tab:disjoin_error}. We also present the percentage of testing data 
that was mismatched following the linear approximation -- that is, the 
percentage of data samples that {\em a posteriori} are observed to be either 
rolling or slipping, but the SVM classifier of the space placed them 
in the wrong category. Furthermore, we present a comparison of the predicted and 
observed data in Figure \ref{fig:fit_goodness}.

\begin{table}[]
	\caption{Error in the fitting of disjoint linear approximations, as 
	described by Equations \eqref{eq:disjointA} and \eqref{eq:disjointB}.}
	\label{tab:disjoin_error}
	\centering
	\begin{tabular}{|ccc|l|ccc|}
		\cline{1-3} \cline{5-7}
		\multicolumn{3}{|c|}{Campaign 
		A}                                                 &  & 
		\multicolumn{3}{c|}{Campaign 
		B}                                                 \\ \cline{2-3} 
		\cline{6-7} 
		\multicolumn{1}{|c|}{}                  & \multicolumn{1}{c|}{Roll}    
		& Slip    &  & \multicolumn{1}{c|}{}                  & 
		\multicolumn{1}{c|}{Roll}    & Slip    \\ \cline{1-3} \cline{5-7}
		\multicolumn{1}{|c|}{Overall error $\delta$}   & 
		\multicolumn{2}{c|}{6.56 
			\%}           &  & \multicolumn{1}{c|}{Overall error $\delta$}   & 
		\multicolumn{2}{c|}{7.84 \%}  \\ \cline{1-3} \cline{5-7}  
		\multicolumn{1}{|c|}{$\delta$}          & \multicolumn{1}{c|}{6.79 \%} 
		& 5.95 \% &  & \multicolumn{1}{c|}{$\delta$}          & 
		\multicolumn{1}{c|}{8.07 \%} & 7.31 \% \\ \cline{1-3} \cline{5-7} 
		\multicolumn{1}{|c|}{$\Delta\,\dot{x}$} & \multicolumn{1}{c|}{4.13 \%} 
		& 3.36 \% &  & \multicolumn{1}{c|}{$\Delta\,\dot{x}$} & 
		\multicolumn{1}{c|}{5.33 \%} & 3.76 \% \\ \cline{1-3} \cline{5-7} 
		\multicolumn{1}{|c|}{$\Delta\,\dot{y}$} & \multicolumn{1}{c|}{2.13 \%} 
		& 2.01 \% &  & \multicolumn{1}{c|}{$\Delta\,\dot{y}$} & 
		\multicolumn{1}{c|}{2.52 \%} & 0.920\% \\ \cline{1-3} \cline{5-7} 
		\multicolumn{1}{|c|}{$\Delta\,\omega$}  & \multicolumn{1}{c|}{3.94 \%} 
		& 3.59 \% &  & \multicolumn{1}{c|}{$\Delta\,\omega$}  & 
		\multicolumn{1}{c|}{4.74 \%} & 5.41\%  \\ \cline{1-3} \cline{5-7} 
		\multicolumn{1}{|c|}{Mismatched data}   & \multicolumn{2}{c|}{7.27 
		\%}           &  & \multicolumn{1}{c|}{Mismatched data}   & 
		\multicolumn{2}{c|}{9.59 \%}  \\ \cline{1-3} \cline{5-7} 
	\end{tabular}
\end{table}

\subsection{Rigid-bounce models}
The parameters of best fit for the rigid bounce model are presented in the first column of Table \ref{tab:penner_fit}, for each of the two campaigns. Also given are the fitting errors
measured using $\delta$, and $\Delta \dot{x}$, $\Delta \dot{y}$ and $\Delta \omega$ as described in the previous section. 

For Penner's extension to the model, the results of the fit are presented in the second two columns of 
Table \ref{tab:penner_fit} for each campaign. 
Observe that the error is slightly smaller than for the simple rigid-body fit, yet still much larger than the SVM fit. Furthermore, the approximation with a 
constant angle of rotation of the frame $\beta$ has slightly smaller error than Penner's proposed
approximation linear relationship \eqref{eq:penner_angle}. 

\begin{table}[]
	\caption{Best fit of parameters from each campaign to rigid bounce models, together with the error. See Methods for details.}
	\label{tab:penner_fit}
	\begin{tabular}{cccclcccc}
		\cline{1-4} \cline{6-9}
		\multicolumn{4}{|c|}{Campaign A}  & \multicolumn{1}{l|}{} & \multicolumn{4}{c|}{Campaign B}          \\
		 \cline{2-4} \cline{7-9} 
		 \multicolumn{1}{|c|}{}   & \multicolumn{1}{c|}{rigid} & 
                 \multicolumn{1}{c|}{varying $\beta$} & \multicolumn{1}{c|}{fixed $\beta$}  &       
		 \multicolumn{1}{|c|}{}   & 	 \multicolumn{1}{|c|}{}   
   & \multicolumn{1}{c|}{rigid} & 
                 \multicolumn{1}{c|}{varying $\beta$} & \multicolumn{1}{c|}{fixed $\beta$}
                 \\  
                \cline{1-4} \cline{6-9} 
		\multicolumn{1}{|c|}{$\beta$}  & \multicolumn{1}{c|}{} 
  & \multicolumn{1}{c|}{$k_P = 
		1.543 \times 10^{-4}$} & \multicolumn{1}{c|}{$\beta = 
		12.9^{\circ}$} & 
		\multicolumn{1}{l|}{} & \multicolumn{1}{c|}{$\beta$}  & \multicolumn{1}{c|}{} & 
		\multicolumn{1}{c|}{$k_P = 2.973 \times 10^{-4}$} & 
		\multicolumn{1}{c|}{$\beta = 18.4^{\circ}$} \\ 
		\cline{1-4} \cline{6-9} 
		\multicolumn{1}{|c|}{$r$}      &  \multicolumn{1}{c|}{0.544} & 
		\multicolumn{1}{c|}{0.448}   &
		 \multicolumn{1}{c|}{0.420}  & \multicolumn{1}{l|}{} & 
		\multicolumn{1}{c|}{$r$}      & \multicolumn{1}{c|}{0.260} & 
		\multicolumn{1}{c|}{0.222} & \multicolumn{1}{c|}{0.147}            \\ \cline{1-4} \cline{6-9} 
		\multicolumn{1}{|c|}{$\mu$}    & \multicolumn{1}{c|}{0.997} &  
		\multicolumn{1}{c|}{0.969}   & 
		 \multicolumn{1}{c|}{0.852}   & \multicolumn{1}{l|}{} & 
		\multicolumn{1}{c|}{$\mu$}    & \multicolumn{1}{c|}{0.998} & 
		\multicolumn{1}{c|}{0.999}                                           &
		 \multicolumn{1}{c|}{0.998}                  \\ \cline{1-4} \cline{6-9} 
		\multicolumn{9}{l}{}        \\ \cline{1-4} \cline{6-9} 
		\multicolumn{1}{|c|}{$\delta$}  &  \multicolumn{1}{c|}{31.9\%} &  		\multicolumn{1}{c|}{25.5\%}  &
		 \multicolumn{1}{c|}{19.0\%}  & \multicolumn{1}{l|}{} & 
		\multicolumn{1}{c|}{$\delta$} & \multicolumn{1}{c|}{35.0\%} & 
		\multicolumn{1}{c|}{21.3\%}  &
		 \multicolumn{1}{c|}{19.2\%}                 \\ \cline{1-4} \cline{6-9} 
		\multicolumn{1}{|c|}{$\Delta\,\dot{x}$} & \multicolumn{1}{c|}{19.3\%} & 
		\multicolumn{1}{c|}{14.6\%}   &
		 \multicolumn{1}{c|}{9.58\%}                 & \multicolumn{1}{l|}{} & 
		\multicolumn{1}{c|}{$\Delta\,\dot{x}$} & \multicolumn{1}{c|}{19.8\%} & 
		\multicolumn{1}{c|}{9.42\%}    &
		 \multicolumn{1}{c|}{9.16\%}                 \\ \cline{1-4} \cline{6-9} 
		\multicolumn{1}{|c|}{$\Delta\,\dot{y}$} & \multicolumn{1}{c|}{4.58\%} & 
		\multicolumn{1}{c|}{4.42\%}  &
		 \multicolumn{1}{c|}{3.17\%}                 & \multicolumn{1}{l|}{} & 
		\multicolumn{1}{c|}{$\Delta\,\dot{y}$} & \multicolumn{1}{c|}{6.32\%}  & 
		\multicolumn{1}{c|}{4.30\%}                                             
		                                                                 &
		 \multicolumn{1}{c|}{4.24\%}                 \\ \cline{1-4} \cline{6-9} 
		\multicolumn{1}{|c|}{$\Delta\,\omega$}  & \multicolumn{1}{c|}{18.6\%} & 
		\multicolumn{1}{c|}{16.1\%}                                             
		                                                                &
		 \multicolumn{1}{c|}{14.2\%}                 & \multicolumn{1}{l|}{} & 
		\multicolumn{1}{c|}{$\Delta\,\omega$}  & \multicolumn{1}{c|}{24.6\%} &
		\multicolumn{1}{c|}{16.4\%}                                                                                                        &
		 \multicolumn{1}{c|}{15.1\%}                 \\ \cline{1-4} \cline{6-9} 
	\end{tabular}
\end{table}

\section{Discussion}

The data presented in this investigation represent the most comprehensive
set we of 
golf ball bounce to data that we are aware of in the open scientific literature. We have taken care to present all steps involved in data collection, curation and error analysis.
Using video analysis and modern image processing techniques, we have 
been able to capture the relationships between the velocity and spin
on lift-off to those on landing, for two different turfs.
We have also made both this summary data and the raw videos
freely available;  our purpose being to enable 
future development and evaluation of 
physically realistic models of golf ball bounce. As a first step in that process, we have evaluated here three simple models; a plain rigid-bounce model with restitution and Coulomb friction, Penner's empirical extension to this model, and a simple piecewise-affine fit to the data. 

We have also been careful to distinguish the prediction interval of the extraction of each quantity from the video, from the natural variability and repeatability of each measurement. Note that while some of the prediction intervals may appear unrealistically large for a few data points, most prediction intervals are around $5\%$ of the mean value of each quantity. Also, in all cases, we took the mean value from these intervals for performing our data analysis, for which the true error is likely to be significantly smaller. 

When looking at the repeatability of the measurements, it is worth noting that in the majority of cases, the confidence intervals on the outbound measurements (horizontal bars in Fig.~\ref{fig:repeatibility}) are  
wider than those on inbound. This indicates that behaviour of the turf was not fully  deterministic; and this points to the variability of turf response. Although present for artificial turf (Campaign A) this effect was less noticeable for artificial  turf (Campaign A), which is as expected. 

The consideration of tangential velocities at landing 
and lift-off enabled us to distinguish between slip and rolling, which appears to be key to
understanding the dynamics of the bounce of the golf ball. Furthermore, we
found little evidence of any ball that lifts off in pure rolling
(see Figure \ref{fig:tang_vel}). Rather, balls that enter rolling during the
bounce enter slip in the opposite sense, prior to take-off. In light of this discovery, we postulate that a thorough consideration of different effects of friction and tangential compliance is missing from current models.

It is striking from our model fitting results that the the piecewise-affine fit outperformed both the rigid-bounce and Penner's models, with 
overall prediction errors around a factor of five times
smaller, for each campaign. As a note of caution, the piecewise-affine model required pre-identification using a SVM of the hyperplane that separates rolling from slipping bounces.  
Note from that the highest contributions to
the error is in the approximation for the outbound horizontal velocity
and spin.  There is evidence that suggests nonlinear behaviour in these
particular variables.  

Furthermore, the SVM fit does not provide any
insight into the physical principles behind the bounce, and the fit yields a discontinuous model.  Not only is there a non-physical
distontinuity in
velocity across the slip/roll boundary values discontinuous, but both fits
are offset from the  origin by a positive value in the $\dot{y}_F$ direction. This would imply that a ball with zero
normal velocity (i.e.~permanently in contact with flat turf) would spontaneously lift off. This observation provides further evidence that a better model
is nonlinear. 
Such nonlinearity could arise due to a dynamic transition elastic behaviour for low normal velocity and elasto-plastic behaviour for higher speed bounces.

For the rigid bounce model, Table \ref{tab:penner_fit} shows that
the average error in predictions for the horizontal velocity and spin
were beyond reasonable approximation (19.3\% and 18.6\%, respectively
for Campaign A; and 19.8\% and 24.6\%, respectively for Campaign B).
The reason for this poor fit is likely to be due to the rigid bounce
model not allowing for any forces other than friction to act in the
horizontal direction.  Most likely, there is an additional
horizontal force due to tangential compliance of the ground. Such
effects are explored in \cite{biber_analysis}.

While providing a slightly better fit than the rigid bounce model, Table \ref{tab:penner_fit} shows that Penner's model is still 
unreasonably inaccurate (14.6\% and 16.1\%, respectively for Campaign A;
and 9.4\% and 16.4\%, respectively for Campaign B) in the
predictions for lift-off horizontal velocity and spin. Note further
that using a fixed constant angle $\beta$ gives a better fit for both
campaigns than Penner's model. Hence, our data does not validate Penner's model when considering our
range of landing conditions. 

Furthermore, Figure \ref{fig:tang_vel} shows 
balls lifting off with reversed tangential velocity; a behaviour not
possible under the rigid-bounce theory. 
Penner's model could qualitatively explain this phenomenon, however we found that no
constant of proportionality in Penner's angle \eqref{eq:penner_angle} applied to the initial conditions classified as rolling was able to map them to a zero value of $v_p$. Nor did we find that any constant angle $\beta$ for that was able to make $v_p \approx 0$ for all such landing conditions.

\section{Conclusion}

We conclude that a reliable physics-based model of golf-ball bounce is still lacking. Such a model would require a complex nonlinear model of the
compliant surface. Most of the discrepancy between the
rigid-bounce models and the data are in the tangential and spin
degrees of freedom, with error in the normal direction being a factor
of five smaller. This observation, along with some of the theoretical
conclusions in \cite{biber_analysis}, suggests the need for a combined frictional and
tangential compliance model of the turf-ball interaction when studying the bounce of a golf ball. Such models would also need to explain the evidence of nonlinearity we have
found in the data.

\section*{Acknowledgements}
 We would like to thank our industrial partner R~\&~A rules Ltd. All experimental work was carried out using R~\&~A's research facilities. We particularly acknowledge helpful advice from Andrew Johnson and Steve Otto. 

\section*{Statements and declarations}
This work was being partially funded by the EPSRC and by the 
industrial partner R~\&~A Rules Ltd.

\noindent The authors have no competing interests to declare that are relevant to the content of this article.

\bibliographystyle{siam} 
\bibliography{refs}

\newpage
\appendix

	\section{Data} \label{sec:data_scatter}

\begin{figure}[h]
	\centering
	\begin{subfigure}{0.95\linewidth}
		\centering
		\includegraphics[width=\linewidth]{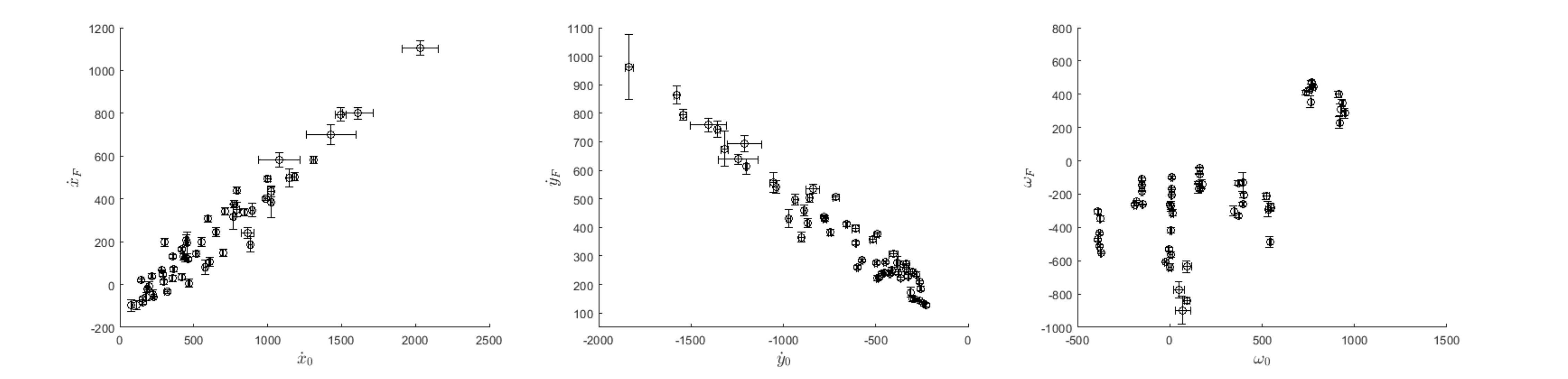}
		\caption{ }
	\end{subfigure}	
	\begin{subfigure}{0.95\linewidth}
		\centering
		\includegraphics[width=\linewidth]{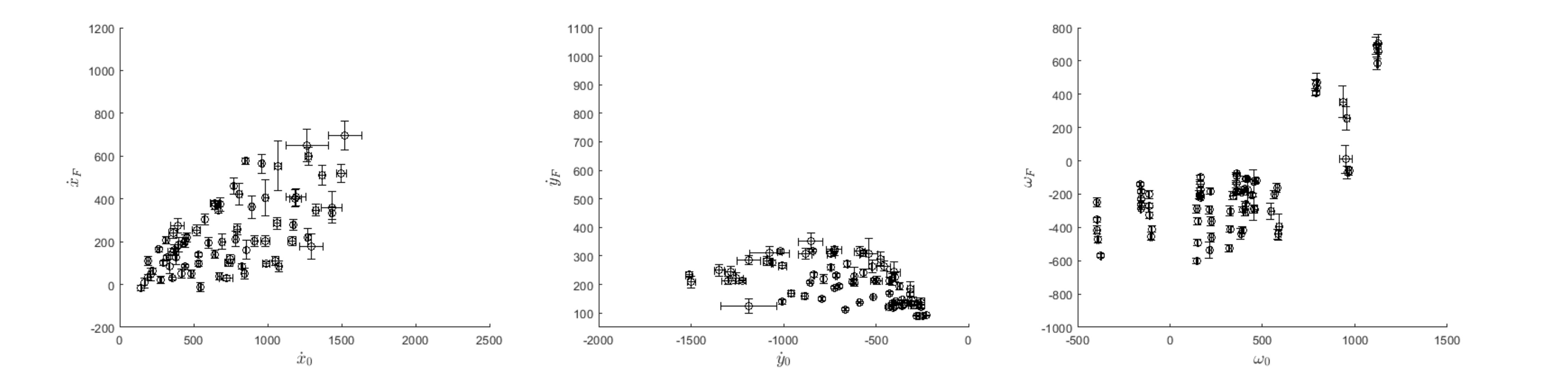}
		\caption{ }
	\end{subfigure}	
	\caption{Mean landing and lift-off values for each setting of the launcher (denoted using circles) together with 95\% CI of the estimates (denoted 
	  using horizontal and vertical error bars). Data is presented for (a) Campaign A, (b) Campaign B; in each case the panels from left to right
          show horizontal velocity, vertical velocity and spin, respectively.}
	\label{fig:repeatibility}
\end{figure}

The mean inbound and outbound speeds and spin measured for each trajectory
are summarised in Figure \ref{fig:repeatibility}, in various two-dimensional slices along with the confidence intervals for each measurement. 

\begin{landscape}

	\begin{figure}[!htb]
		\centering
		\begin{subfigure}{0.47\linewidth}
			\includegraphics[width=0.95\linewidth]{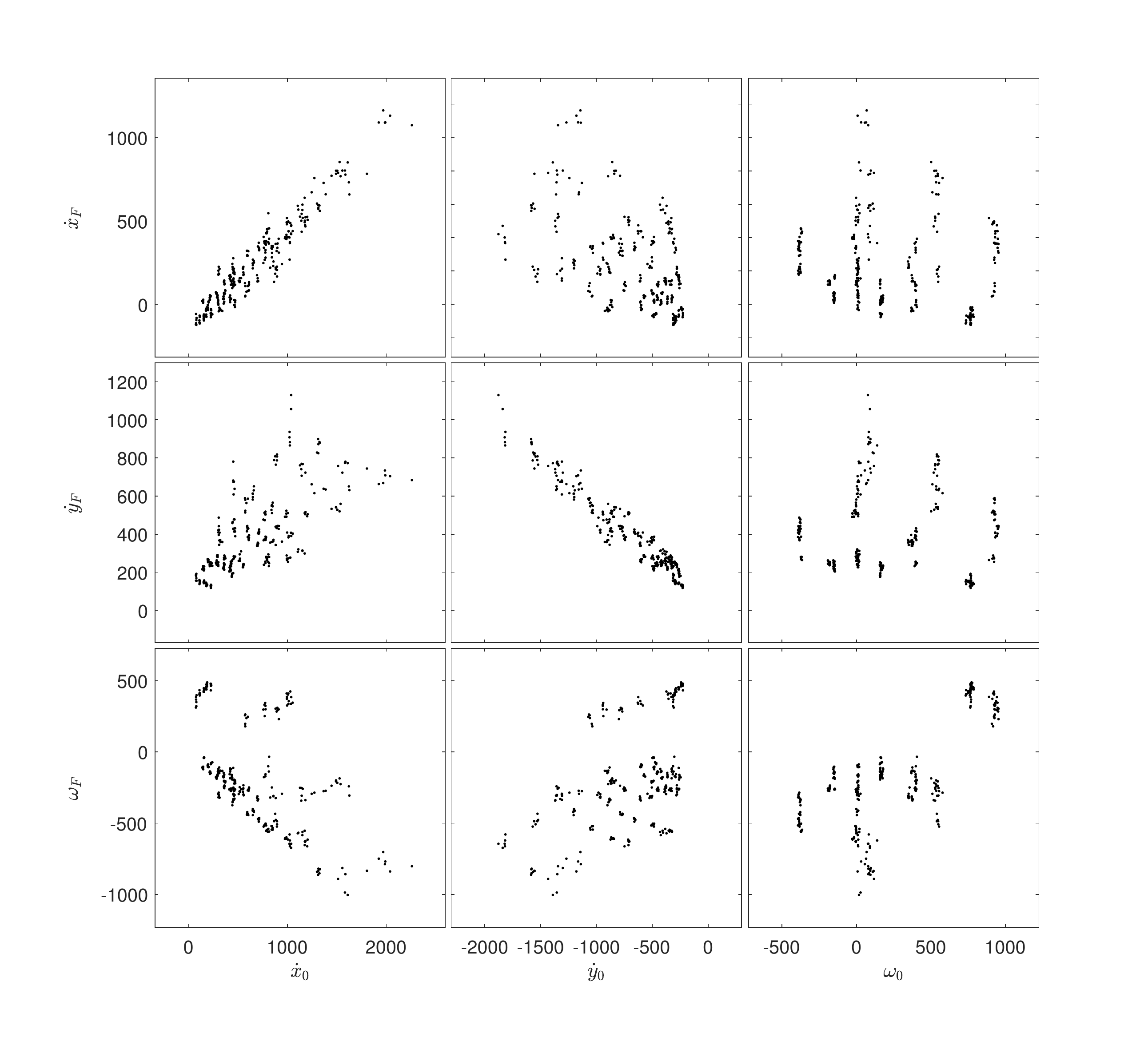}
			\caption{ }
		\end{subfigure}
		\begin{subfigure}{0.47\linewidth}
			\includegraphics[width=0.95\linewidth]{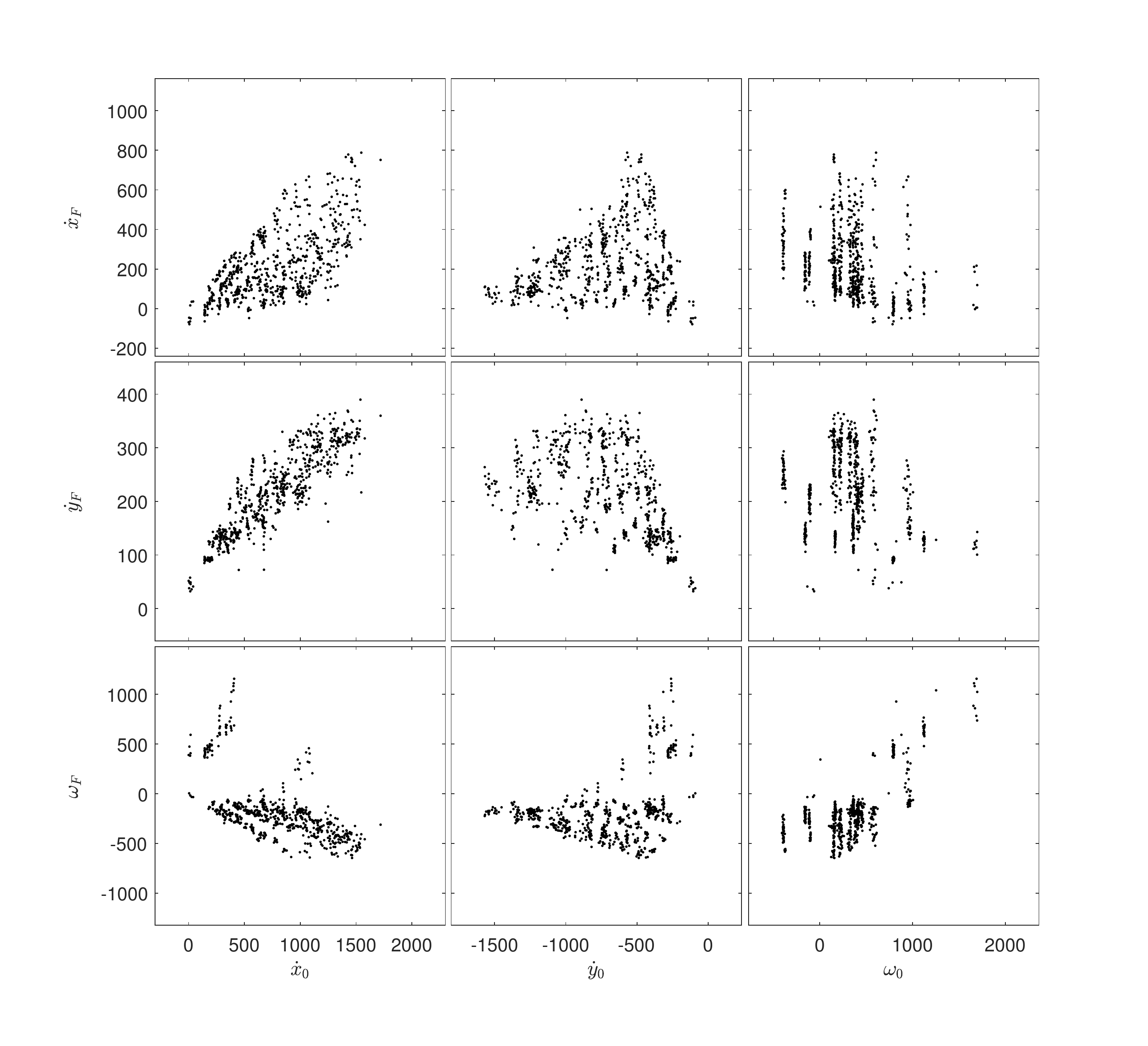}
			\caption{ }
		\end{subfigure}
		\caption{A full representation of all available data from (a) Campaign 
		A; (b) Campaign B. Here, lift off values are presented as the outcome 
		of varying the inbound quantities. }
		\label{fig:data_scatter}
	\end{figure}
	
\end{landscape}

The complete data for the same quantities (without means and error bars) are presented in all possible two-dimensional slices in illustrated in Figure \ref{fig:data_scatter}. The complete data set, and the videos from which they were obtained can be freely downloaded from \cite{videos}. 

\newpage 

\end{document}

%% file: macros.tex
\usepackage[utf8]{inputenc}
\usepackage[margin = 1in]{geometry}
\usepackage{amsmath}
\usepackage{amsfonts}
\usepackage{amssymb}
\usepackage{overpic}
\usepackage{epstopdf}
\usepackage{graphicx}
\usepackage{bm}
\usepackage{xcolor}
\usepackage[font = small]{caption}
\usepackage[font = small]{subcaption}
\usepackage{authblk}
\usepackage{lscape}
\usepackage{url}

\renewcommand{\vec}[1]{\ensuremath{\bm{#1}}}

\newcommand{\ve}{\ensuremath{\varepsilon}}

\newcommand{\ga}{\ensuremath{\alpha}}